\documentclass[a4paper,12pt]{article}
\pdfoutput=1 
\usepackage{graphicx}
\usepackage{jheppub} 
\usepackage{tikz,tkz-euclide}
\newcommand{\R}{{\mathbb R}}
\newcommand{\C}{{\mathbb C}}
\newcommand{\be}{\begin{equation}}
\newcommand{\eeq}{\end{equation}}
\newcommand{\bea}{\begin{eqnarray}}
\newcommand{\eea}{\end{eqnarray}}
\newcommand{\ba}{\begin{array}}
\newcommand{\ea}{\end{array}}
\def\nn{\nonumber}
\newcommand{\ft}[2]{{\textstyle\frac{#1}{#2}}}

\newcommand{\ii}{\mathrm{i}}
\newcommand{\ee}{\end{equation} }

\newcommand{\one}{{\rm 1\kern -.9mm l}}

\title{ Wilson Loops and Chiral Correlators on  Squashed Spheres}

\author[a]{F. Fucito, J.F. Morales}
\author[b]{and R.Poghossian}

\affiliation[a]{I.N.F.N - sezione di Roma 2\\
and Universit\`a di Roma Tor Vergata, Dipartimento di Fisica\\
Via della Ricerca Scientifica, I-00133 Roma, Italy}
\affiliation[b]{Yerevan Physics Institute,\\
Alikhanian Br. 2, AM-0036 Yerevan, Armenia}
\emailAdd{fucito@roma2.infn.it}
\emailAdd{morales@roma2.infn.it}
\emailAdd{poghos@yerphi.am}

\abstract{  We study chiral deformations of  ${\cal N}=2$ and ${\cal N}=4$ supersymmetric gauge theories
obtained by turning on $\tau_J \,{\rm tr} \, \Phi^J$ interactions  with $\Phi$ the ${\cal N}=2$ superfield.
 Using localization, we compute the deformed gauge theory partition function $Z(\vec\tau|q)$
 and the expectation value of circular Wilson loops $W$ on a squashed four-sphere.
 In the case of the deformed ${\cal N}=4$ theory, exact formulas for $Z$ and $W$ are derived  in terms of an underlying $U(N)$  interacting matrix model replacing the free Gaussian model describing the ${\cal N}=4$ theory.
 Using the AGT correspondence, the $\tau_J$-deformations are related to the insertions of commuting integrals of motion in the four-point CFT correlator and chiral correlators are expressed as $\tau$-derivatives of the gauge theory partition function on a finite $\Omega$-background.   In the so called Nekrasov-Shatashvili limit, the entire ring of chiral relations is extracted from the $\epsilon$-deformed Seiberg-Witten curve.
As a byproduct of our analysis we show that $SU(2)$ gauge theories on rational $\Omega$-backgrounds are dual to CFT minimal models.

}

\keywords{SYM theories, Wilson loops, recursion relation, large N, AGT.}

\preprint{ROM2F/2015/04}
\begin{document}

\maketitle
\flushbottom
\section{Introduction and summary}

Wilson loops provide some of  the most interesting observables to investigate
in  a gauge theory. Many of the intrinsic features of the gauge dynamics are encoded by Wilson loops of specific shapes. The quark-antiquark potential
in a gauge theory with matter can be extracted from the asymptotic fall off  of a Wilson loop made of two straight lines. The cusp anomalous dimension of
the gauge theory  can be extracted from the behaviour of the Wilson loop near a cusp.  In the context of the AdS/CFT correspondence, the Wilson loop at strong coupling computes the minimal area on AdS of a string worldsheet  ending on the loop at the AdS boundary and provides some of the most remarkable tests of the correspondence.

In theories with extended supersymmetry, the natural loops to consider are those preserving a fraction of the supersymmetry. The supersymmetric Wilson loop
 typically  involves both the gauge field $A_m$ and
its scalar superpartners  $\varphi_i$
\be
W=\langle {\rm tr} \,e^{\cal C} \rangle  \qquad {\rm with} \qquad {\cal C}=\ii \int (A_m \dot x^m+\varphi_i \, \dot y^i) ds
\ee
 Supersymmetry requires $|\dot x|=|\dot y|$. Many such solutions have been found preserving different fractions of the original  supersymmetry \cite{Drukker:1999zq,Zarembo:2002an,Drukker:2007qr}.

In this paper we  focus on circular Wilson loops for theories with ${\cal N}=2,4$ supersymmetry on a squashed $S^4$.
Circular Wilson loops in ${\cal N}=4$ theory  were first considered in \cite{Erickson:2000af} where it  was conjectured that the perturbative series for $W$ at large N  reduces to a simple counting of ladder diagrams.  The counting of diagrams was extracted from a Gaussian matrix model \cite{Brezin:1977sv}.
 An exact  formula for $W$ was then proposed in \cite{Drukker:2000rr}  by an explicit evaluation of the Gaussian
matrix model partition function to all orders in $1/N$.

The conjectured formula for the  circular Wilson loop in the ${\cal N}=4$ theory was later proved in \cite{Pestun:2007rz}  using  localization. 
 In that reference, the partition function of a general ${\cal N}=2$  gauge theory  on a round sphere $S^4$ was shown to be given by the integral $\int da |Z(a,\tau)|^2$ with $Z(a,\tau)$ the gauge partition function on $\R^4$. Here $Z$ is evaluated  on an $\Omega$-background such that $\epsilon_1=\epsilon_2=\epsilon$ are finite, so the gauge theory lives effectively on a non-trivial
gravitational background \cite{Nekrasov:2003rj,Billo:2006jm}.  The function $Z(a,\tau)$ was previously computed in \cite{Nekrasov:2002qd,Flume:2002az,Bruzzo:2002xf,Nekrasov:2003rj,Nakajima:2003uh,Losev:2003py} by localization in the moduli space of instantons for theories with various gauge groups and matter content.  On the other hand the expectation value of a circular Wilson loop on $S^4$ was shown to be given by the same integral with the insertion of
  the phase ${\rm tr} \,e^{2\pi i a\over \epsilon}$ inside the integral.  Both $W$ and $Z$  are corrected by instantons and anti-instantons localized around the north and south poles of the sphere, whose contributions can be most easily accounted for if they come from the two different patches of which our space time is made of.  The ${\cal N}=4$ Gaussian matrix model was then recovered from  the case of ${\cal N}=2^*$ theory in the limit where the mass of the adjoint hypermultiplet is sent to zero.

  It is natural to ask how these results are modified when the YM action is deformed or the spacetime is changed. An interesting spacetime to consider is that
  of a squashed sphere \cite{Hama:2012bg}. A motivation for considering four-dimensional gauge theories on squashed spheres comes from the AGT correspondence that relates the partition function of ${\cal N}=2$ supersymmetric theories to correlators in a dual two-dimensional CFT with the squeezing
parameter  $\epsilon_1/\epsilon_2$ parametrizing the central charge of the CFT  \cite{Gaiotto:2009we,Alday:2009aq,Drukker:2009id}.
  In \cite{Alday:2009fs}, it was proposed that the expectation value of a circular Wilson loop oriented along, let us say the first plane, on the squashed sphere is given by the
  insertion of  the phase ${\rm tr} \,e^{2\pi i a\over \epsilon_1}$ in the partition function integral.
  In this paper we will prove this formula using localization and extend it to the case of a general ${\cal N}=2$ theory with prepotential
  ${\cal F}_{\rm class}=\tau_J\, {\rm tr}\,\Phi^J$.
  This class of gauge theories were first introduced in \cite{Losev:2003py}  and the deformed partition function  $Z(a,\vec\tau)$  was computed  in \cite{Nakajima:2003uh} (see also \cite{Nekrasov:2009rc}) and  related to the generating function of Gromov-Witten invariants for certain complex manifolds.

   In this paper we present a "physical derivation" of the deformed partition function  based on localization and apply the results to the study of circular Wilson loops on squashed spheres. The crucial observation is that the circular Wilson loop ${\cal C}$  is nothing but the lowest component of the  equivariant superfield
   ${\cal F}$ defining the field strength of a connection on  the so called universal moduli space, the moduli space of instantons times the space time. Indeed in presence of an $\Omega$ background the zero mode solutions to the equations of motion get deformed and the lowest component of ${\cal F}$ reduces to
 $\tilde\varphi =\varphi+\delta_\xi x^m\, A_m$ with $\delta_\xi x^m$ a rotation of the spacetime coordinates along the Cartan of the Lorentz group.  The same combination appears on the circular Wilson loop  ${\cal C}={2\pi \ii n_1 \over \epsilon_1} \tilde\varphi $, so the expectation value of the Wilson loop
 can be related to the  $\Omega$-deformed version of the corresponding chiral correlator. Moreover using the fact that Wilson loops are allowed only on spheres with  rational squeezing parameter $\epsilon_1/\epsilon_2$, an explicit evaluation of  $\tilde\varphi$ shows that
 $\left\langle  e^{\cal C} \right\rangle =\left\langle e^{ {2\pi \ii n_1 \over \epsilon_1} a} \right\rangle $, so all instanton corrections to  ${\cal C}$ (but not to the correlator) cancel after exponentiation !  The same cancellation was observed in \cite{Pestun:2007rz} for the case of the round sphere.
   Similarly ${\rm tr}\,\Phi^J$ interactions can be related to the higher component of ${\rm tr}\, {\cal F}^J$,  and their expectation values are again given by the
   deformed chiral correlators up to some super-volume factors.

    The case of the deformed ${\cal N}=4$ theory is particularly interesting. $\tau_J$-interactions break supersymmetry down
    to ${\cal N}=2$  but surprisingly the resulting theory is far more simpler than any other ${\cal N}=2$ theory one can think of.
       For instance, as we will see, instanton contributions to the partition function and circular Wilson loops can be shown to cancel in
       the deformed theory more in the same way than in the  maximally supersymmetric theory. As a consequence the gauge partition function and the expectation values of circular Wilson loops  are described by an effective interacting matrix model.
  More precisely, the Wilson loop is given by the integral
    \be
    \left\langle \, {\rm tr} \, e^{ \cal C} \right\rangle ={1\over Z} \int_\R  d^Na\, \Delta(a)  \, {\rm tr} \,e^{ 2\pi  a \over \epsilon_1}\, e^{- N\,  V(\vec\tau,a)}
    \ee
with $\Delta(a)$ the Vandermonde $U(N)$ determinant  and $V(\vec\tau,a)$ a potential codifying the $\tau_J$-interactions.
The matrix model integral effectively counts the number of Feynman diagrams with propagators ending on the Wilson loop, and internal
J-point vertices weighted by $\tau_J$.  The case with only quartic interactions will be worked out
in full details. A formula for $W$ at large
$N$ but exact in $g_{YM}$ will be derived in this case. This formula provides a simple prediction $S=\ln W \sim \sqrt{\lambda_{\rm eff}}$  for  the minimal area  of a string worldsheet  on the gravity dual theory
as a function of a $\tau_4$-dependent effective t'Hooft coupling $\lambda_{\rm eff}$. As a bonus, in this case, it will also be possible to find the correlators of the Wilson loop with ${\rm tr}\,\Phi^2$ and 
${\rm tr}\,\Phi^4$ and check our results against those of \cite{Semenoff:2001xp}\footnote{We thank the referee of the paper to point out this reference to us.}

    Exploiting the AGT correspondence between 4d gauge theories and 2d CFT's,  the chiral correlators $\langle \, {\rm tr} \, \tilde\varphi^J \rangle$ in the gauge theory can be related to insertions of the
integrals of motions $I_J$ into the four-point CFT correlator computing the gauge theory partition function \cite{Nekrasov:2009rc,Bonelli:2014iza}.   The computation of such deformed correlators is feasible and it leads to relations which are the analogue
of the chiral ring relations for ${\cal N}=2$ gauge theories
\cite{Cachazo:2002ry} for a finite $\Omega$ background. We carry out the explicit computation in the $SU(2)$ theory with four fundamentals
for the simplest chiral correlators and show the match with the dual  correlators in the Liouville theory.
We also consider higher point correlators in the so called Nekrasov-Shatshvili limit in which one of the two deformation parameters of the $\Omega$ background is sent to zero. In this limit an
$\epsilon$-deformed version of the Seiberg-Witten type curve is available \cite{Poghossian:2010pn,Fucito:2011pn} from which we extract the full set of chiral ring relations.

Finally, focusing on rational $\Omega$-backgrounds (those admitting Wilson loops) we propose a duality between certain $SU(2)$ gauge theories with
four fundamental hypermultiplets of critical masses  and minimal
models (see \cite{Santachiara:2010bt,Estienne:2011qk,Bershtein:2014qma,Alkalaev:2014sma,Belavin:2015ria} for previous studies of this duality).
These critical gauge theories fall into a general class of models studied in  \cite{Fucito:2013fba} where the entire instanton contributions
can be explicitly resummed in terms of hypergeometric functions. The gauge duals of all  4-point correlators
of the Ising models are identified and the gauge partition functions are matched against the dual CFT correlators.

    The paper is organised as follows: In section 2 we review the derivation of the deformed partition function for ${\cal N}=2$ gauge theories with adjoint or
    fundamental matter on $\R^4$. In section 3 we consider the gauge theory on the squashed sphere $S^4$.  We derive a localization formula
    for the Wilson loop in the deformed theory and discuss in details the case of the ${\cal N}=4$ theory deformed by a quartic $\tau_4$-interaction. In section 4
     we describe the  AGT duality between the chiral correlators in the $SU(2)$ gauge theory and the insertions of integrals of motion in the Liouville four-point function.
  We show that chiral correlators in the gauge theory can be written in terms of derivatives of the gauge theory partition function and show how chiral
  ring relations can be extracted from  the $\epsilon$-deformed Seiberg-Witten  curve  in the limit
in which one of the two parameters of the $\Omega$-background is sent to zero. In section 5,  we describe the duality between $SU(2)$ gauge theories
with critical masses and rational $\Omega$-backgrounds and CFT minimal models.

\section{The gauge partition function on $\R^4$}

The action of a general ${\cal N}=2$  supersymmetric gauge theory on $\R^4$ is specified by a single holomorphic function,  {\it the prepotential},
${\cal F}_{\rm class}(\Phi)$  of the ${\cal N}=2$ vector multiplet superfield $\Phi$. Explicitly
 \be
S_{\rm class}=\left[ {1\over 4 \pi^2} \int d^4x d^4 \theta \, {\cal F}_{\rm class}(\Phi)+{\rm h.c.}\right]+\ldots
\ee
 with $(x_{m},\theta_{\alpha\dot a})$ denoting the super-space coordinates and the lower dots denoting couplings to hypermultiplets. We are interested in the Coulomb branch of the gauge theory where
the scalar $\varphi$ in the vector multiplet $\Phi$ has a non-trivial vacuum expectation value.
In this branch, only the vector multiplet prepotential is corrected by quantum effects.
 We take the classical prepotential of the general form
\be
{\cal F}_{\rm class}(\Phi)= \sum_{J=2}^p  {2\pi  \ii \tau_J\over J!} \, {\rm tr}\,   \Phi^J   \label{fphi}
\ee
for some integer $p$ and
\bea
\Phi &=& \varphi+ \lambda_m \theta^m+ \ft12 \,F_{mn} \theta^m \theta^n +\ldots    \label{phisuper}
\eea
  Here and in the rest of the paper we used the "twisted" fermionic variables $\theta^m=\ft12 \sigma^{m\alpha\dot a} \theta_{\alpha\dot a}$ obtained by identifying
 internal  $\dot a$ and Lorenz  $\dot \alpha$ spinors indices.
 The gauge theory following from (\ref{fphi}) can be seen as a deformation of the standard renormalizable gauge theory
where the gauge coupling is replaced by a function of the scalar field  $\varphi$ and the super symmetrically related  couplings $F\lambda^2$ and $\lambda^4$ are included.
The standard theory with prepotential  ${\cal F}_{\rm class}(\Phi)= \pi \ii \tau  {\rm tr}\,   \Phi^2$ is recovered after setting  $\tau_{J>2} $ to zero.

We will refer to the partition function $Z(\vec \tau ,q)$ as the ``deformed partition function".
 There are two main contributions to the partition function $Z_{\rm one-loop}$ and $Z_{\rm tree+inst}(\vec \tau,q)$ coming from the fluctuations
 of the gauge theory fields around the trivial and the instanton vacua.  We notice that only the  latter one depends
 on the couplings $\tau_J$ since  $Z_{\rm one-loop}$ is given by a one-loop vacuum amplitude.  On the other hand $Z_{\rm tree+inst}(\vec \tau,q)$
 can be written in terms of an integral over the instanton moduli space that can be computed with the help of localization.
 We start by reviewing the construction of the instanton moduli space and introducing the basic ingredient for localization: the equivariant supercharge $Q_\xi$
 and the equivariant vector superfield ${\cal F}$.  For concreteness we focus on conformal gauge theories with gauge group $U(N)$  and fundamental or adjoint matter.

\subsection{The instanton moduli space and the equivariant charge}

The moduli space of instanton solutions can be packed into
 a bosonic-fermionic pair of $[N+2k]\times 2k$  matrices $\Delta_0, {\cal M}$
 characterizing the instanton gauge connection  and the gaugino zero modes.  The instanton gauge
 connection is built out of the ADHM matrix $\Delta=\Delta_0 + x_m b_m$, with $b_m$ a constant
 and moduli independent matrix.
 The matrices $\Delta, {\cal M}$  satisfy the super-ADHM constraints
 \be
\bar{\Delta}     \, \Delta \sim
  \one_{[2]\times [2]}      \qquad \bar \Delta\, {\cal M}+\bar{\cal M}\, \Delta=0 \label{adhm0}
\ee
  The ADHM constraints (\ref{adhm0}) together with the obvious $U(k)$ symmetry of the construction lead to $4k^2$ bosonic and fermionic relations for the $4 kN+4k^2$ bosonic and fermionic degrees of freedom in $\Delta_0$,
  ${\cal M}$ leaving a $4kN$-dimensional  superspace $\mathfrak M_{k,N}$.Fundamental
 hypermultiplets  contribute  an extra $[k\times N_f]$ matrix ${\cal K}$ of fermionic moduli that can be conveniently paired with the auxiliary field $h$ in the same representation.
The instanton connection can be written in terms of the [N+2k]$\times$ [N]  matrix $U$ defined by
\be
\Delta U=\bar U \Delta=0  \qquad ~~~~~~~~~~~  \bar U\, U=\one
\ee
 and reads
  \be
 A_m=\bar U \,\partial_m  U
 \ee
    To localize the integral  one first introduces the equivariant charge on $\mathfrak{M}\times \C^2$,
    \be
    Q_\xi=Q+d+i_\xi
    \ee
    with $d=dx^m \partial_m $ the exterior differential on $\C^2$, $Q$ a  supersymmetry component  acting on the moduli space via
  \bea
  Q \, \Delta &=& {\cal M}    \qquad    ~~~~~~~~~   Q\,  {\cal M}=0   \nn\\
      Q\,  {\cal K} &=& h    \qquad  ~~~~~~~~~~~~Q \, h = 0
  \eea
 and  $i_\xi$ a contraction defined by
 \be
 i_\xi \, dx^m=\delta_\xi x^m \qquad ~~~~~~~   i_\xi \, d\mathfrak M_i=\delta_\xi \mathfrak M_i
 \ee
with $\delta_\xi$ an element of the symmetry group $SO(4)\times U(k)\times U(N)\times U(N_f)$ rotating moduli and spacetime coordinates.
 The combined action can then be written as
   \bea
  Q_\xi  \, \Delta &=& {\cal M} +d\Delta   \qquad  ~~~~~~     Q^2_\xi\,  \Delta=\delta_\xi \Delta \nn\\
    Q_\xi  \, {\cal K}  &=&  h \qquad    ~~~~~~~~~~~~~~~Q^2_\xi \,{\cal K}  = \delta_\xi  {\cal K}
  \eea
 We notice that $Q^2_\xi =\delta_\xi$, so $Q_\xi$ is nilpotent at the fixed points of  $\delta_\xi$.
 The charge $Q_\xi$ can be thought of as an equivariant version of the supersymmetry charge. The second ingredient for localization, is the
 construction of an equivariant field ${\cal F}$ defined by
\be
{\cal F}= Q_\xi( \bar U \, Q_\xi U) +[ \bar U Q_\xi U, \bar U Q_\xi U] =\tilde \varphi  +\lambda+F
\ee
with
 \bea
\tilde \varphi &=&  \bar U\, \delta_\xi U  + Q\bar U  \,Q U+[ \bar U Q U, \bar U Q U]    \nn\\
\lambda &=& \nabla (\bar U \, Q U)  \nn\\
   F &=& \nabla ( \bar U d U)
 \eea
the $0$, $1$ and $2$-form parts of ${\cal F}$ on $\C^2$ and $\nabla=d+ [A,.]$.  We notice that at finite $\epsilon$'s the field $\tilde\varphi$ involves a non-trivial linear combination of the scalar
$\varphi$ and the vector field $A_m$ components.  Indeed, recalling that  $\delta_\xi $ rotates both the instanton moduli and the
  spacetime variables $x^m$ one can write \cite{Flume:2004rp}
 \be
  \tilde \varphi= \bar U\, \delta^{\rm mod}_\xi U  + Q\bar U  \,Q U+ [ \bar U Q U, \bar U Q U]+  \delta_\xi x^m \, A_m=\varphi   +  \delta_\xi x^m \, A_m\label{phitilde}
 \ee
In the limit $\epsilon_\ell$ to zero,  $\delta_\xi x^m  \to 0$,  $\tilde{\varphi}$ reduces to $ \varphi$ and the equivariant correlator
computes the scalar correlators of the gauge theory on $\R^4$.   In this paper, we are interested on
   gauge theories living on a non-trivial $\Omega$-background so we  keep $\epsilon_1,\epsilon_2$ finite and the supersymmetric
   operators are  built out of $\tilde\varphi$ rather than $\varphi$.

It is important to notice that ${\cal F}$ coincides with the ${\cal N}=2$ vector superfield $\Phi$ (\ref{phisuper})
after replacing  $\theta^m$ by $dx^m$ and evaluating the fields on the instanton background. Indeed, one can check that higher $\theta$-terms
in (\ref{phisuper}) cancel on the instanton solution.
The field   ${\cal F}$ can then be viewed as the equivariant version of the ${\cal N}=2$ superfield $\Phi$.
Moreover
  \be
 Q_\xi \, {\cal F} = (\delta_\xi+\delta_{\rm gauge}) (\bar U \,Q_\xi U) \label{inv}
  \ee
  i.e. ${\cal F}$ is $Q_\xi$-invariant up to a symmetry rotation. This implies that any invariant function made out of ${\cal F}$ is $Q_\xi$-closed.
  In particular the Yang-Mills action (\ref{fphi}) in the instanton background  can be written in the explicit  $Q_\xi$-invariant form
  \be
   S_{\rm inst}(\vec \tau) = \sum_{J=2}^p {\ii \tau_J\over 2\pi J!} \,  \int_{\C^2} {\rm tr}\, {\cal F}^J     \label{fj}
   \ee

\subsection{The deformed gauge partition function}

The gauge partition function on $\R^4$ can be written  as the product of the one-loop, tree level and instanton contributions
    \be
     Z(\vec \tau)=   Z_{\rm one-loop}  \,  Z_{\rm inst+tree}(\vec\tau)  \label{ztotal0}
     \ee
      The one-loop partition function $Z_{\rm one-loop}$   is given by
    \be
Z_{\rm one-loop }=  Z^{\rm gauge}_{\rm one-loop }\,  Z^{\rm matter}_{\rm one-loop }   \label{zoneloop0}
\ee
with \cite{Alday:2009aq,Fucito:2013fba}
\bea
Z^{\rm gauge}_{\rm one-loop } &=&  \prod_{u < v}^N      \Gamma_2(a_{uv} )^{-1}  \Gamma_2(a_{uv}+\epsilon )^{-1}      \nn\\
Z^{\rm matter}_{\rm one-loop } &=&  \left\{
\begin{array}{ll}
 \prod_{u<v}^N   \Gamma_2(a_{uv}-m)  \Gamma_2(a_{uv}+m+\epsilon)       &   {\rm adj.}   \\
  \prod_{u,v=1}^N  \Gamma_2( a_v-\bar{m}_{u}  ) \Gamma_2( a_u-m_{v+N}+\epsilon)   & {\rm fund.}    \\
\end{array}
\right.
   \label{zloop}
 \eea
 Here $\Gamma_2(x)$ is the Barnes  double gamma function\footnote{  The  Barnes  double gamma function can be thought of as a regularization of the infinite product (for $\epsilon_1>0,\epsilon_2>0$)
 \be
\Gamma_2(x) =\prod_{i,j=0}^\infty   \left( {\Lambda \over x+i \epsilon_1+j\epsilon_2}\right)
\ee
} and $\epsilon=\epsilon_1+\epsilon_2$.

 The instanton partition function is defined by the moduli space integral
   \be
 Z_{\rm inst+tree}( \tau) =  \sum_{k=0}^\infty   \, \int d\mathfrak M_{k,N} \,e^{-S_{\rm inst}( \vec\tau) }    \label{zinst}
 \ee
 with the integral running over the instanton moduli space for a given $k$. The action $S_{\rm inst}$ and the integral over the instanton moduli
 spaces can be evaluated with the help of localization.  According to the localization theorem, given an equivariant derivative $Q_\xi$, and an
 action $\delta_\xi=Q^2_\xi$ on a space $M$,  the integral of any $Q_\xi$-closed equivariant form $\alpha$ is given by  the localization formula \cite{Bruzzo:2002xf}
   \be
   \int_{M} \alpha=(-2\pi)^{\frac{D}{2}} \sum_{x_0}    {\alpha_0\over {\rm det} \delta_\xi}   \label{locform}
   \ee
   with $D$ the complex dimension of the space $M$,  $x_0$ labelling the fixed points of $\delta_\xi$ and $\alpha_0$ the 0-form part, with respect to the equivariant $Q$-grading, of $\alpha$.
 To evaluate $S_{\rm inst}(\vec \tau)$, we take $M=\C^2$, $\alpha=\sum_{J=2}^\infty {\tau_J\over 2\pi J!} \,   {\rm tr}\, {\cal F}^J $ and
  $\delta_\xi$ a rotation in the Cartan of the SO(4) Lorentz group acting  as  $z_\ell \to e^{i \epsilon_\ell } \, z_\ell$ on the complex coordinates of $\C^2$.
   The  unique fixed point of this action is the origin and the localization formula leads to
\be
S_{\rm inst}(\vec \tau)=  \sum_{J=2}^p  {\ii \tau_J\over 2 \pi J!}  \, \int_{\C^2} {\rm tr}\, {\cal F}^J     =
{2\pi  \ii \over  \epsilon_1 \epsilon_2}  \, \sum_{J=2}^p  {\tau_J\over J!}  \,  {\rm tr}\, {\tilde \varphi}(0)^J    \label{sphitilde}
   \ee
    To perform the integral over the instanton moduli space we take  $\alpha=e^{-S_{\rm inst}(\vec\tau)}$  and
     $\delta_\xi$ an  element of the Cartan subgroup of  the symmetry group $ U(N)\times U(N_f)\times U(k)\times SO(4)$. We  parametrize
     the Cartan element by $a_u, m_i, \chi_I, \epsilon_{1,2}$. More precisely, $a_u$ parametrizes the eigenvalues of the vacuum expectation value
     matrix $\langle \varphi \rangle$, $m_i$ the masses of the fundamental or adjoint fields, $\chi_I$ the
     Cartan of U(k) and  $\epsilon_{1,2}$ are Lorentz breaking parameters that deform the
    $\R^4$ spacetime geometry. For a gauge theory on flat space $\epsilon_{1,2}$ should be sent to zero at the end of the computation. For finite $\epsilon$'s the integral describes the partition function
     on a non-trivial gravitational  background, the so called $\Omega$-bakcground. Fixed points of $\delta_\xi$ are again isolated
   and are  in one-to-one correspondence with N-tuples of Young tableaux $Y=(Y_1,\ldots Y_N)$ with a total number of $k$ boxes. The Young tableaux $Y$
   specify the U(k) Cartan elements $\chi_I$. Explicitly
   \be
   \chi_{(i,j)}=a_u+(i-1) \epsilon_1+(j-1) \epsilon_2      \qquad   (i,j)\in Y_u    \label{chiij}
   \ee
   In our conventions $\epsilon_1,\epsilon_2$ are pure real numbers.
  From (\ref{locform}) one finds for the moduli space integral
    \be
    Z_{ \rm inst+tree}( \tau) =\sum_{Y}    Z_Y\, e^{-S_Y(\vec \tau)}   \label{zint}
\ee
with $Z_Y$ the inverse determinant of $\delta_\xi$ and $S_Y(\vec \tau)$ the 0-form of the instanton action (\ref{sphitilde}) evaluated at the fixed point.
 For $Z_Y$ one finds \cite{Flume:2002az,Bruzzo:2002xf,Fucito:2013fba}
    \be
Z_Y= { 1 \over {\rm det} \, \delta_\xi }\Big|_{Y}=   Z_Y^{\rm gauge} \, Z_Y^{\rm matter}
\ee
with
\bea
Z_Y^{\rm gauge} &=& \prod_{u,v}^N    Z_{Y_u,Y_v}(a_{uv})^{-1}    \nn\\
Z_Y^{\rm matter} &=&   \left\{
\begin{array}{ll}
   \prod_{u,v}^N   Z_{Y_u,Y_v}(a_{uv}+m)     &   {\rm adjoint}    \\
  \prod_{u,v=1}^N    Z_{\emptyset,Y_v}( \bar{m}_{u}-a_v  ) \, Z_{Y_u,\emptyset}(a_u-m_{v+N} ) &  {\rm fund}    \\
\end{array}
\right.
  \label{zinst0}
\eea
and
\bea
  Z_{Y_u,Y_v}(x)  &=& \prod_{ (i,j)\in Y_u}
      (x-\epsilon_1(k_{vj}-i)+\epsilon_2 \, (1+\tilde k_{ui}-j)) \nn\\
      \times && \prod_{ (i,j)\in Y_v} (x+\epsilon_1(1+ k_{uj}-i)-\epsilon_2 \, (\tilde k_{vi}-j))
      \label{zalbe2}
 \eea
 Here $(i,j)$ run over  rows and columns respectively of  the given Young tableaux,
$\{ k_{uj} \} $  and $ \{ \tilde k_{ui} \}$ are positive integers
giving the length of the rows and columns respectively of the tableau $Y_u$.
 We remark that  (\ref{zloop}) and (\ref{zinst0}) are not
symmetric under the exchange  of fundamental $\bar m_i$ and  anti-fundamental masses $\bar m_j$ but a totally symmetric form
under the exchange $\bar m_i \leftrightarrow \bar m_j$ can be obtained by replacing
 \be
m_{u+N} \to  \bar m_{u+N}+\epsilon
 \ee
   Finally, the contribution of the deformed Yang-Mill action  at the fixed point $Y$ reduces to
  \be
S_Y(\vec \tau) = {2\pi \ii \over \epsilon_1 \epsilon_2}  \, \sum_{J=2}^p  {\tau_J\over J!} \,    {\rm tr}\, {\tilde \varphi}_0^J\, \big|_{Y}
= {2\pi \ii \over \epsilon_1 \epsilon_2}  \, \sum_{J=2}^p  {\tau_J\over J!} \,   {\cal O}_{J,Y}
\ee
    with $ \tilde\varphi_0=\bar U \, \delta_\xi U$
   the zero form of the equivariant superfield $\cal F$ and ${\cal O}_{J,Y}$ the chiral primary operator evaluated at the fixed point. 
     To evaluate ${\cal O}_{J,Y}$, one first notices that for a $\Delta$ specified by the Young tableaux data $Y$
   one can find  infinitely many solutions $ U^{(k,l)}$ to the defining equation $\bar  \Delta U=0$. The $\delta_\xi$-eigenvalue
of $U^{(k,l)}$ is given by $  \chi_{(k,l)}=a_u+(k-1)\epsilon_1+\epsilon_2(l-1)$ with $(k,l)\notin Y_u $  \cite{Flume:2004rp}.
 Summing over $(k,l)$ one finds for the generating function of the chiral operators  \cite{Losev:2003py,Flume:2004rp}.
 \bea
&&  {\rm tr}\, e^{ z {\tilde \varphi}_0} \, \big|_{Y} ={\cal V} \sum_{u=1}^N\sum_{(k,l) \notin Y_u}  e^{z  \chi_{(k,l)} }  =
 {\cal V} \sum_{u=1}^N  \left( \sum_{k,l } e^{z  \chi_{(k,l)} } -\sum_{k,l \in Y_u}e^{z  \chi_{(k,l)} } \right)   \nn\\
  &&=  \sum_u\left( e^{z a_u}
 - (1-e^{z\, \epsilon_1})(1-e^{z\, \epsilon_2})   \sum_{(i,j)\in Y_u} e^{z \chi_{(i,j)} } \right)  \label{varphifix0}
\eea
  where  we have included the normalization factor ${\cal V}=(1-e^{z\, \epsilon_1})(1-e^{z\, \epsilon_2})  $  in order to reproduce the classical
 result $\langle e^{z \varphi} \rangle _{\rm class}={\rm tr} \, e^{z a}$. Expanding both sides of (\ref{varphifix0}) in powers of $z$ one  finds   
 \be
{\cal O}_{J,Y}=   \sum_u\left( a_u^J
 -   \sum_{(i,j)\in Y_u}    \left[ \chi_{(i,j)}^J+(\chi_{(i,j)}+\epsilon)^J-  (\chi_{(i,j)}+\epsilon_1)^J-(\chi_{(i,j)}+\epsilon_2)^J    \right]  \right)  
\label{varphifix}
\ee
 For the first few values of $J$ one finds
 \bea
  {\cal O}_{2,Y}&=&{\rm tr} a^2 - 2\,k\, \epsilon_1\,\epsilon_2\nn\\
 {\cal O}_{3,Y}&=&{\rm tr} a^3 - 3\, \epsilon_1\,\epsilon_2\, \sum_{u=1}^N  \sum_{(i,j)\in Y_u} (  \epsilon_1+\epsilon_2+2\chi_{(i,j)}   )\nn\\
  {\cal O}_{4,Y}&=&{\rm tr} a^4 - 4\, \epsilon_1\,\epsilon_2\, \sum_{u=1}^N  \sum_{(i,j)\in Y_u} ( 2 \epsilon_1^2+3\epsilon_1\, \epsilon_2+ 2 \epsilon_2^2
  +6\,  \epsilon\,\chi_{(i,j)}  + 6\,\chi_{(i,j)}^2   )  \label{ojs}
 \eea
 The deformed partition functions can then be written as
    \be
   Z_{\rm inst+tree} (\vec \tau)=  \sum_{Y}    Z_Y(\vec \tau) =  \sum_{Y}    Z_Y\,  {\rm exp}   \left( - {2\pi \ii \over \epsilon_1 \epsilon_2}  \, \sum_{J=2}^p {\tau_J\over J!} \,    {\cal O}_{J,Y}  \right)   \label{zdef}
    \ee
    with   ${\cal O}_{J,Y} $ given by (\ref{varphifix}). The gauge prepotential ${\cal F}_{\rm eff}$
 is then identified with the free energy associated to the deformed instanton partition function
  \be
 {\cal F}_{\rm eff} ( \vec \tau )=-   \epsilon_1 \, \epsilon_2\, \ln Z(\vec \tau)   \label{fdef}
 \ee
      We  notice that the deformed partition function (\ref{ztotal0}) and the prepotential (\ref{fdef})  reduce  to the partition function and prepotential
    of the undeformed ${\cal N}=2$ gauge theory if we take $\tau_{J>2}=0$.
     Indeed using (\ref{ojs}) for $J=2$ one finds
    \be
    Z( \tau_J=\tau \delta_{J,2})=   e^{-{ \pi \ii \tau \sum_u a_u^2 \over \epsilon_1 \epsilon_2}} \, Z_{\rm oneloop} \, \sum_{Y}    Z_Y \, q^{|Y|}
    \ee
    with $q=e^{2\pi \ii \tau}$ and $Z_{\rm tree}=  e^{-{ \pi \ii \tau \sum_u a_u^2 \over \epsilon_1 \epsilon_2}}  $ the tree level contribution to the partition
    function and  $Z_{\rm inst}=\sum_{Y}    Z_Y \, q^{|Y|}$ the instanton part.

 \subsection{Chiral correlators  }

In presence of a non-trivial $\Omega$-background, the $\epsilon$-deformed chiral correlator $\langle  {\rm tr}\,  \tilde\varphi^J (x) \rangle $
provides a simple class of supersymmetric observables of the gauge theory.  Supersymmetric Ward identities show that these correlators are
independent of the position $x$.

It is easy to see that the chiral correlators can be extracted from the deformed partition function. To this aim, we first introduce the $Q_\xi$-invariant volume form
 \be
 \alpha_4=  dz_1\, d\bar z_1\, dz_2\, d\bar z_2\,
 -i (\epsilon_1 |z_1|^2 dz_2 d\bar z_2+\epsilon_2 |z_2|^2 dz_1 d\bar z_1)- \epsilon_1 \epsilon_2  \,z_1\, \bar z_1\, z_2\, \bar z_2\,
 \ee
 and write the chiral operator in the equivariant form
 \be
 \int d^4x\,  {\rm tr}\,  \tilde\varphi^J   =  \int_{\C^2}  \alpha_{4}\wedge   \,  {\rm tr}\,  {\cal F}^J
   \label{chiloc}
 \ee
 Using the localization formula (\ref{locform}) one finds for  the normalized chiral correlator
\be
 \langle  {\rm tr}\,  \tilde\varphi^J  \rangle =   { \left\langle   \int_{\C^2}  \alpha_{4}\wedge   \,  {\rm tr}\,  {\cal F}^J \right \rangle \over \left\langle   \int_{\C^2} \alpha_4  \right\rangle }
 =  {1\over Z_{\rm inst+tree}  } \sum_Y Z_Y(\vec\tau)\,   {\cal O}_{J,Y}
 \ee
  or equivalently from (\ref{zdef})
  \be
 {1\over J!}  \langle  {\rm tr}\,  \tilde\varphi^J \rangle
 ={\ii \epsilon_1 \epsilon_2 \over 2\pi } \, \partial_{\tau_J} \ln Z(\vec{\tau}) \label{phij}
 \ee
  Formula  (\ref{phij}) generalizes  the so called Matone relation that relates
   $ \langle  {\rm tr}\,  \varphi^2  \rangle$ to the $\tau$-derivative of the prepotential  in the undeformed theory in flat space.
   On the other hand, it shows that one can view the deformed partition function as the generating functional of the
   general multi-trace chiral correlators
   \be
    \langle  {\rm tr}\,  \tilde\varphi^{J_1} \, {\rm tr}\,  \tilde\varphi^{J_2} \,\ldots  \rangle_{\rm undeformed}
   \ee
   in the undeformed theory.

 \section{The gauge theory on $S^4$ }

  The gauge partition function on $S^4$ is given by the integral  \cite{Pestun:2007rz}
    \be
 Z_{S^4} (\vec \tau)  =c\,   \int_{\gamma}  d^N a \,   |  Z_{\rm one-loop}(  a )\,  Z_{\rm tree+inst} ( a ,\vec \tau) |^2\label{ZS4}
 \ee
 with $d^N a=\prod_u d a_u $ {}\footnote{Notice that  in our conventions the Vandermonde determinant is reabsorbed into the one-loop
 determinant $ Z_{\rm one-loop}$. } and $c$ a normalization constant. The integral runs along the imaginary axis.
The partition functions  $Z_{\rm one-loop}$ and $Z_{\rm inst}$ are given by (\ref{zloop}) and (\ref{zinst0}) with vevs and masses taken
in the domains
\be
a_u =\in \ii \R  \qquad   \qquad   m,\epsilon_\ell \in \R   \qquad   m_u={\epsilon\over 2}+ \ii \R   \qquad    \bar{m}_u=-{\epsilon\over 2} + \ii \R 
\ee
These domains are chosen such that complex conjugate of  the one-loop partition function (\ref{zloop}) is given by the same formula with
$\Gamma_2(x)$ replaced by $\Gamma_2(\epsilon-x)$.
We  adopt units where $\epsilon_1\epsilon_2=1$ and write $\epsilon_1=\epsilon_2^{-1}=b$. In these units the one-loop partition function  becomes
\bea
|Z^{\rm gauge}_{\rm one-loop }|^2 &=&  \prod_{u < v}^N      \Upsilon( a_{uv} )  \Upsilon(-a_{uv} )      \nn\\
|Z^{\rm matter}_{\rm one-loop }|^2 &=&  \left\{
\begin{array}{ll}
 \prod_{u,v}^N      \Upsilon(  a_{uv}-m)^{-1}     &   {\rm adj.}   \\
  \prod_{u,v=1}^N  \Upsilon( a_v-\bar{m}_{u}  )^{-1} \Upsilon(m_{v+N}- a_u)^{-1}   & {\rm fund.}    \\
\end{array}
\right.
   \label{zloops4}
 \eea
 with
\be
\Upsilon(x)={1\over \Gamma_2\left(x|b,\ft{1}{b}\right)\Gamma_2\left(b+\ft{1}{b} -x|b,\ft{1}{b}\right)  }
\ee
 $\Upsilon$ is an entire function  satisfying $\Upsilon(x)=\Upsilon(\epsilon-x)$. It has an infinite number of single zeros at $x=-m b-n/b$ and $x=(m+1)b+(n+1)b$
 for $m,n\geq 0$ integers.
  Finally the normalization  $c$ has been fixed for later convenience to be\footnote{This normalization is chosen in such a way that the partition function and its AGT dual
  correlator precisely match. }
  \be
  c=q^{ {1\over 2} m_3(m_3+\epsilon)+ {1\over 2} m_4(m_4+\epsilon) }  \label{ccc}
  \ee

  \subsection{ Wilson loops  }

  A supersymmetric Wilson loop  is defined by the line integral
   \be
{\cal C}= \ii \int_0^L (A_m \,\dot{x}^m + |\dot{x}| \, \varphi_1 ) ds  \label{cc}
   \ee
  with  $\varphi_1=\ft12(\varphi-\varphi^\dagger)$  and $A_m$ taken to be anti-hermitian matrices.
  We use complex coordinates $x^m=(z_1,z_2,\bar z_1, \bar z_2)$ and consider a circular Wilson loop
  defined by the path
   \be
  z_\ell (s)= r_\ell \, e^{ \ii \epsilon_\ell s}     \label{solc}
  \ee
  with
   \be
  L= {2\pi  n_1\over \epsilon_1}={2\pi n_2\over \epsilon_2} \label{epsratio}
   \ee
  The condition (\ref{epsratio}) ensures that the path is closed and it can be satisfied if and only if
  the ratio $\epsilon_1/\epsilon_2$ is rational.
  Moreover, taking $r_{1,2}$ satisfying
   \be
   |\dot x|^2= \epsilon_1^2 |r_1|^2+ \epsilon_2^2|r_2|^2=1
  \label{solc1}
  \ee
   one finds
   \be
  \dot x^m =( \ii \epsilon_1 z_1 ,  \ii \epsilon_2 z_2,  -\ii \epsilon_1 \bar z_1 ,  -\ii \epsilon_2 \bar z_2)= \delta_\xi x^m
   \ee
and the Wilson loop can be written in the suggestive form
        \be
    {\cal C}= \ii \int_0^L \left( A_m \, \delta_\xi  x^m +   \varphi_1 \right) ds  =\ft{\ii}{2} \int_0^L \tilde{\varphi}(s) ds+{\rm h.c.}
    \ee
    The appearance of the combination $\tilde \varphi$ in the Wilson loop is not surprising, since this is the supersymmetric (equivariant) version of the vector field
    one form $A$. The evaluation of the Wilson loop expectation value can then be performed again with the help of localization by inserting
    in the partition function the operator ${\rm tr} \, e^{\cal C}$. We notice that the 0-form part ${\cal C}_0=\ii \tilde{\varphi}_{0}=\ii \bar U\delta_\xi U$ is anti-hermitian
and that correlators of $\tilde{\varphi}(s)$ do not depend on $s$, so the Wilson loop operator on the instanton background $Y$ reduces to
     \be
   e^{{\cal C}_0}\Big|_Y=  \ {\rm tr}\, e^{ \ii \, L\,  \tilde\varphi_0    } \, \big|_{Y} =\sum_u e^{  {2 \pi \ii \, n_1\over \epsilon_1} \, a_u }
      \ee
      In deriving this relation we used (\ref{varphifix0}) evaluated at  $z={2 \pi \ii \, n_1\over \epsilon_1}  $.  Remarkably, for this special value of $z$, the whole dependence
 of ${\cal C}$ on the instanton configuration $Y$ cancels upon exponentiation!
     The expectation value of the Wilson loop is then given by the simple formula
    \be
  \left \langle {\rm tr}\,   e^{{ \cal C} } \right\rangle_{S^4} =  {1\over Z}
   \int_\gamma d^Na\, {\rm tr} \, e^{2\pi \ii n_1\, a\over \epsilon_1 } \,   \left|   Z_{\rm one-loop} ( a)Z_{\rm tree+inst}( a,\vec \tau) \,     \right|^2    \label{ws40}
 \ee
   For the undeformed theory, formula (\ref{ws40}) was proposed and motivated in \cite{Alday:2009fs}  as the generalization of the more familiar result
   on the round sphere to the case of a squashed sphere with a Wilson loop oriented along the $z_1$-plane.  This loop corresponds to the choice $r_2=0$ in (\ref{solc}). Our result provides a proof
of that formula and a generalisation to the case of ${\cal N}=2$ theories with
  a classical prepotential  of general type. We notice that  $\tau_J$-corrections to the Wilson loop expectation value are computed by the insertions of chiral operators in the Wilson loop correlator, so that
the expectation values of the Wilson loop in the deformed theory can be thought of as the generating function
   of the correlators
   \be
    \left \langle {\rm tr}\,   e^{{ \cal C} } \,   {\rm tr}\,  \tilde\varphi^{J_1}\,\tilde\varphi^{J_2}\, \ldots \right\rangle_{S^4,\rm undef.}
  \label{ws4def}
 \ee
 in the undeformed theory. 
  For example, the first $\tau_J$-correction  to the Wilson loop expectation value is computed by the correlator
 \be
  {\partial \over \partial \tau_J} \left \langle {\rm tr}\,   e^{ \cal C }  \right\rangle_{S^4}   = {2 \pi \ii \over \epsilon_1 \epsilon_2 \, J!}   
  \left \langle {\rm tr}\,   e^{ \cal C } \,   {\rm tr}\,  \tilde\varphi^{J}  \right\rangle_{S^4,\rm undef.}   \label{deltaw}
 \ee

 \subsubsection{Perturbative expansion}

  The correlators  (\ref{ws4def}, \ref{deltaw}) can be computed order by order in perturbation theory in the weak coupling regime ${\rm Im} \tau \to \infty$.
  In this limit, the integral (\ref{ws40}) is dominated by the region where $a$ is small. Expanding the integrand in (\ref{ws40}) in powers of $a$, one can
  compute each term in the expansion in terms of a correlator in an effective matrix model. Using $\Upsilon(x)\sim x$ for $x$ small, and the fact that
  instanton corrections are suppressed in this limit one finds
 \be
 |Z_{\rm oneloop}|^2\sim \Delta(a)=\prod_{u<v} a_{uv}^2 +\ldots \qquad~~~~~~  Z_{\rm inst}=1+\ldots  \label{zonepert}
 \ee
Plugging (\ref{zonepert}) into (\ref{deltaw}) one finds, for the leading $\tau_J$-correction  to the Wilson loop expectation value at weak coupling, the result
 \be
 {\partial \over \partial \tau_J} \left \langle {\rm tr}\,   e^{ \cal C }  \right\rangle_{S^4}  =  {2\pi \ii \over \epsilon_1 \epsilon_2\, J!}      
   \int_\gamma  d^{N} a \, \Delta(a)\, e^{ {2\pi\over \epsilon_1 \epsilon_2}  {\rm Im}\, \tau  \, {\rm tr} a^2}  {\rm tr } \, a^{J}    \,   {\rm tr } \, a^{2}    +\ldots   \label{matrix2}
 \ee
 Similarly, higher order corrections in perturbation theory are given by Gaussian matrix model integrals of type (\ref{matrix2}) involving higher order monomials in $a$.  A direct test of these formulae by an explicit Feynman diagram computation would be very welcome.

 \subsection{The  ${\cal N}=4$ deformed theory}

 In this section we consider the effect of turning on $\tau_J$-interactions on the simplest theory at our disposal: the ${\cal N}=4$ theory.  $\tau_J$-deformations break ${\cal N}=4$ supersymmetry
down to ${\cal N}=2$ by including self-interactions for one of
the three chiral multiplets. The resulting theory is surprisingly simple and, as we will see, it exhibits a perfect  cancellation of instanton contributions to the gauge partition  function.   This is in contrast with the  case of ${\cal N}=2^*$ theory where the ${\cal N}=4$ symmetry is broken by giving mass to
the adjoint hypermultiplet spoiling the  balance between the instanton corrections coming from gauge and matter multiplets.

As in the undeformed case, the ${\cal N}=4$ deformed theory will be defined as a
the limit of the ${\cal N}=2^*$ deformed theory where the mass of the
adjoint hypermultiplet is sent to ``zero" \cite{Pestun:2007rz}.  More precisely, the
points where the ${\cal N}=4$ is restored will be identified with the zeros of the instanton partition function in the $m$-plane. It is easy to see that they are located at
 $m=-\epsilon_1$ or $m=-\epsilon_2$. Indeed,  for any choice of $\tau_J$ and
instanton number the instanton partition function (\ref{zinst0}) can be seen to be always proportional to $(m+\epsilon_1)(m+\epsilon_2)$ with the two factors coming from  the
contributions  to $Z_{Y_u Y_u}(m) $ of the  Young tableaux boxes $(i,j)=(k_{uj},\tilde k_{ui} )\in Y_u$.
 We notice that these two points on the $m$-plane coincide in the case of the round sphere where $\epsilon_1=\epsilon_2=\ft{\epsilon}{2}$.
 and the symmetric point $m=-\ft{\epsilon}{2}$ is unique ($m=0$ in the conventions of \cite{Pestun:2007rz}). For concreteness here we take $m=-\epsilon_1$.

The one-loop partition function (\ref{zoneloop0}) also drastically simplifies  at $m=-\epsilon_1$.
Indeed using the double Gamma function identity
\bea
\Gamma_2(x+\epsilon_1) \,\Gamma_2(x+\epsilon_2)=x\, \Gamma_2(x)\, \Gamma_2(x+\epsilon)
\label{id}
\eea
 one finds
  \be
  |Z_{\rm oneloop}|^2 =\Delta(a)=\prod_{u\neq v} a_{uv}
 \ee
 with $\Delta(a)$ the Vandermonde determinant describing the $U(N)$ measure.
 The gauge partition function reduces to the $U(N)$ matrix model integrals
  \be
  Z=  \int d^Na \,\Delta(a) \, e^{- N\,  V(a,\vec\tau)}    \label{zws4}
 \ee
 with the integral over $a_u$ now running along the real line and the potential defined by 
 \be
 V(a,\vec\tau)=    {2\pi \ii  \over \epsilon_1 \epsilon_2\, N}  \, \sum_{J=2}^p   {\tau_{J}   \over J!} \,   {\rm tr} (\ii a)^J+{\rm h.c.}
 \label{v1}
 \ee
On the other hand the expectation value of a circular Wilson loop is given by
 \be
 W ={1\over Z} \int d^Na \,\Delta(a) \, {\rm tr} \,e^{2\pi  a \over \epsilon_1}\, e^{- N\,  V(a,\vec\tau)}    \label{ws4}
\ee
 We notice that unlike in the ${\cal N}=4$ theory, in presence of $\tau_J$-interactions the matrix model underlying the theory is no-longer
 Gaussian but interacting. The integrals (\ref{zws4}) and (\ref{ws4}) count now  diagrams involving not only propagators but also $J$-point vertices.
 Luckily enough, the underlying matrix models have been extensively studied in the literature.
    In the large $N$ limit integrals (\ref{zws4}) and (\ref{ws4}) have been explicitly evaluated by saddle point methods \cite{Brezin:1977sv}. One defines the resolvent
  \be
  w(x)={1\over N} \left\langle {\rm tr}\, {1\over x-a} \right\rangle={1\over N} \left\langle \, \sum_{u=1}^N {1\over x-a_u} \right\rangle
  \ee
   A simple algebra shows that $w(x)$ defined like this satisfies a quadratic equation with solution
     \be
     w(x)=\ft12\left(  V'(x)-\sqrt{V'(x)-4 f_{p-2}(x) } \right) \label{wxg}
     \ee
where $f_{p-2}(x)$ a polynomial of order $p-2$ determined by the condition that $w(x)\approx {1\over x}$ at large $|x|$.
 We notice that $w(x)$ has a discontinuity along the cuts defined by the zeroes of the square root, so we can write
 \be
w(x\pm \ii 0) =  \int_{\cal S} {\rho(y) dy\over x-y\pm \ii 0}=\ft12 \, V'(x) \pm \pi \, \ii\, \rho(x)
\ee
with $\cal S$ the union of the cuts and $\rho(x)$ the density
 \be
 \rho(x)={1\over N}\sum_u \delta(x-a_u)
 \ee
  It is often enough to assume the presence of a single cut and look for $w(x)$ in the form
   \be
   w(x)=\ft12 V'(x) -Q_{p-2}(x) \sqrt{ (x-b_1)(x-b_2) }  \label{wq}
   \ee
   with $Q_{p-2}(x)$ a polynomial of order $p-2$.  Indeed the number of unknown variables in $\{ Q_{p-2} (x), b_1, b_2 \}$ is $p+1$, matching
   the number of equations coming from requiring that $w(x)\approx {1\over x}$ for large $|x|$ and therefore $w(x)$ is fully determined.

\begin{figure}
\includegraphics[scale=1,height=8cm, width=14cm]{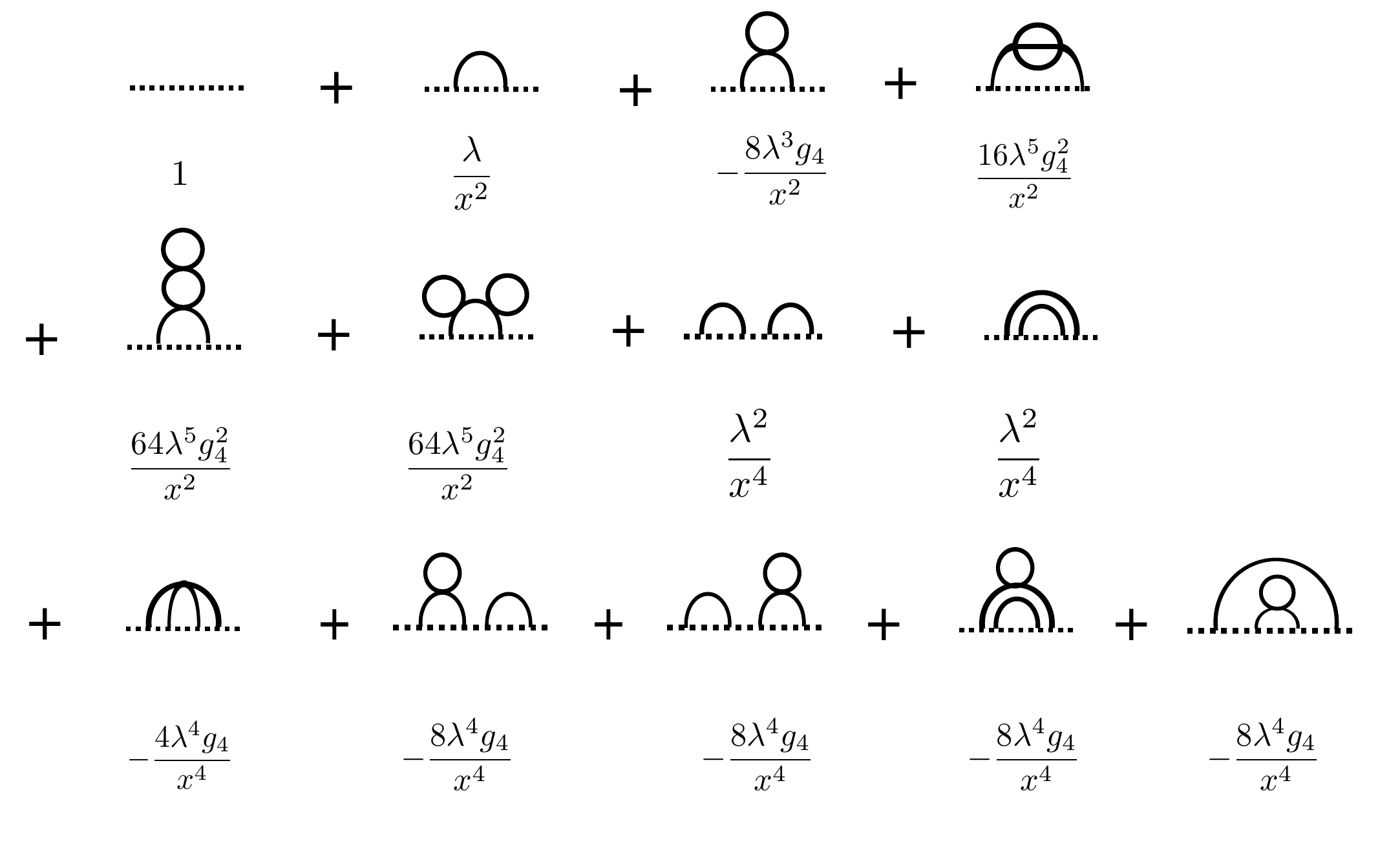}
\caption{Diagrams contributing to  the circular Wilson loops. The dashed line represents the $\tau$-line where Wilson loop operators ${\cal C}(\tau)$ are inserted. }
\label{figure}
\end{figure}

   As an example, let us consider the quartic potential
   \be
   V(a)={1\over 2\lambda} \, a^2+g_4 \, a^4  \label{v2}
   \ee
with
\be
\lambda={ N\epsilon_1\epsilon_2 \over 4\pi ({\rm Im}\tau_2)} ={g^2_{YM}N\epsilon_1\epsilon_2 \over 16\pi^2}  \qquad ~~~~~~ g_4=-{ 4\pi {\rm Im}(\tau_4)\over 4! N\,\epsilon_1\epsilon_2}   \label{lll}
\ee
Since $V(a)=V(-a)$ one can take $b_2=-b_1=2b$ and $Q(x)=c_0+c_2 x^2$ in (\ref{wq}).
     Requiring $w(x)\approx {1\over x}$ for large $|x|$ one finds
   \be
   w(x)={1\over 2\lambda} \left(  x+4\, x^3\, g_4 \, \lambda-(1+4\,\lambda \,g_4  (x^2+2 b^2))\sqrt{x^2-4 b^2}     \right)  \label{wx}
   \ee
   with
   \be
   b^2={ \sqrt{ 1+48 \lambda^2 g_4}-1 \over 24 \lambda g_4}  \label{bsol}
   \ee
   Plugging (\ref{bsol}) into (\ref{wx}) and expanding for large $x$ and small $\lambda$ one finds \cite{Brezin:1977sv}
  \be
x\,  w(x)={1\over N} \sum_{n=0}^\infty \left\langle \, {{\rm tr} \, a^{2n}\over x^{2n} } \right\rangle =1+\frac{\lambda-8 \,
{\lambda}^{3} \, {g}_{4}+\hbox{144} \, {\lambda}^{5} \, {g}_{4}^{2}}{{x}^{2}}+\frac{2 \, {\lambda}^{2}-36 \, {\lambda}^{4} \, {g}_{4}}{{x}^{4}}+\dots
  \label{wxn}
  \ee
 Notice that only the even powers ${\rm tr} \, a^{2n}$ have a non-trivial expectation value as expected for an even potential.
  Terms proportional to $x^{-2n} \, (-g_4)^k\, \lambda^{p}$ count the number of
  diagrams with  $2n$ external lines, $p$ propagators and $k$  four-point vertices. In Fig.\ref{figure} we show the relevant diagrams and contributions leading to (\ref{wxn}).
The entire sum can be written in the  analytic form \cite{Brezin:1977sv}
  \be
  x\, w(x)=1+\sum_{n=1}^\infty\sum_{k=0}^\infty {  (-12 g_4 \lambda^2)^{k}\, \lambda^n\, (2n)! \,(2k+n-1)! \over x^{2n} \, n!\, (n-1)!\, k!\, (k+n+1)! }
  \ee
  For the Wilson loop one then  finds
  \bea
   W  &=& {1\over N} \sum_{n=0}^\infty \left( {2\pi  n_1 \over \epsilon_1}\right)^{2n} \left\langle \, {\rm tr} \, {a^{2n} \over  (2n)! }  \right\rangle\nn\\
 &=& 1+ \sum_{n=1}^\infty \sum_{k=0}^\infty {  (-12 g_4 \lambda^2)^{k}\, \left( {4\pi^2 n_1^2 \lambda \over \epsilon_1^2 } \right)^n\,  (2k+n-1)! \over n! \,(n-1)!\, k!\, (k+n+1)! }\nn\\
 &=&    {2\, \over\sqrt{ \lambda_{\rm eff} }} \left[   I_1 \left(\sqrt{ \lambda_{\rm eff} } \right)  + {   \delta \,   \over   \delta+2 } I_3\left(\sqrt{ \lambda_{\rm eff} } \right)  \right]
     \label{wilsonfin}
 \eea
with
  \be
   \lambda_{\rm eff} =    {32\pi^2 n_1^2 \lambda \over \epsilon_1^2  (\delta+2)   }
   =  {2 \, g_{YM}^2  N \, n_1 n_2  \over   (\delta+2)  }     \quad \quad \delta=\sqrt{1+48 \, g_{4} \, \lambda^2 } -1    \label{leff}
      \ee
   Formula (\ref{wilsonfin}) gives the large $N$ limit of the Wilson loop expectation value in the $\tau_4$-deformed ${\cal N}=4$ theory and it is exact in $\lambda_{\rm eff}$.
  When  $g_4 \approx 0$ and $n_1 n_2=1$ one finds the familiar ${\cal N}=4$
  formula   \cite{Drukker:1999zq,Erickson:2000af,Drukker:2000rr} 
   \be
  W\approx {2\, I_1 \left(\sqrt{  \lambda_{\rm YM} } \right)     \over \sqrt{   \lambda_{\rm YM}  }    }    \qquad  {\rm with}
  \qquad   \lambda_{\rm YM} = g_{\rm YM}^2 N 
   \ee
   On the other hand in the limit of large $\lambda_{\rm eff}$ with $g_4 \lambda^2$ kept finite  one finds
  \be
  W  \approx  {4\,e^{\sqrt{\lambda_{\rm eff} } }  \over \sqrt{2\pi} \lambda_{\rm eff}^{3\over 4} }  \left({ \delta+1 \over \delta+2 }\right)
  \ee
    This result  provides us with a remarkable simple prediction for the AdS dual of the deformed theory. Indeed according to holography, in the limit of large $N$ and $\lambda_{\rm eff}$,
   the area of a string world sheet on AdS ending on the loop is given by $\ln W$ so  one expects
   \be
    S =\ln  W   \approx  \sqrt{\lambda_{\rm eff}} +\ldots
    \ee
    On the gravity side, a general class of solutions preserving half of the supersymmetries of $AdS_5\times S^5$ were constructed  in \cite{D'Hoker:2007fq}.
    These solutions are specified by a single function $w(x)$ with branch cuts in the x-plane.  It is natural to identify the gravity dual of the deformed ${\cal N}=4$ theory with the one-cut solution specified by the matrix model resolvent $w(x)$. The minimal string world sheet area
   in the $w$-geometry can be computed  along the lines of  \cite{D'Hoker:2007fq} providing a precise test of the duality.

\subsection{ Two point correlators $ \langle {\rm tr} e^{\cal C} \,{\rm tr} \varphi^{J} \rangle$  }

 The expectation value of the Wilson loop in the deformed theory can be viewed as the generating function of correlators in the undeformed 
 gauge theory involving the insertions of chiral primary operators in the Wilson loop. These correlators compute the expansion coefficients of chiral primary operators in the OPE of the 
 circular Wilson at distances much larger than the size of the loop.  In \cite{Semenoff:2001xp},  correlators 
 $ \langle {\rm tr} e^{\cal C} \,  {\rm tr} \varphi^{J} \rangle$ computing the leading coefficients in the expansion were evaluated  for the ${\cal N}=4$ theory on $\R^4$.  In this section we show how these results 
for $J=2,4$ can be extracted from our localization formulae.
 
   The crucial observation in \cite{Semenoff:2001xp} is that the two point function $\langle  A_M \dot x^M (x) A_N \dot x^N  (y) \rangle $ with $x^M(s)$ the ten-dimensional path along the circular loop does 
not depend on the insertion point positions. Moreover, the contributions to the
   correlators $ \langle {\rm tr} e^{\cal C} \, {\rm tr} \varphi^{J} \rangle$ coming from Feynman diagrams involving internal vertex insertions are argued to cancel to all orders in perturbation theory.  
The correlator is then effectively computed by a Gaussian matrix model with the insertion of a J-point vertex counting the number or
   rainbow diagrams ending on the loop with the extra J-point insertion. The same matrix model computes the leading $\tau_J$-correction of the 
   deformed ${\cal N}=4$ gauge theory on $S^4$ in the limit where all $\tau_J$'s are small. The results for the correlators for 
   $J=2,4$ can then be extracted from the expansion of the exact formula (\ref{wilsonfin}) in the limit where $\tau_2\approx \tau$ and $\tau_4\approx 0$
   i.e. the undeformed theory.  
\vskip .5cm
\begin{center}
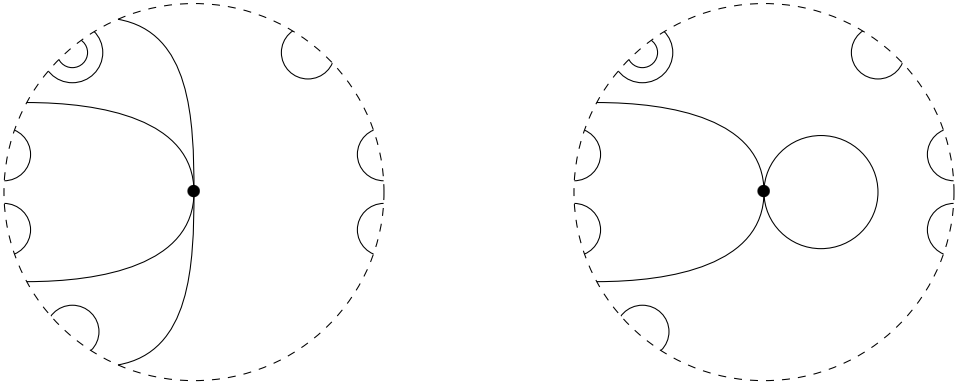
\begin{figure}
\begin{tikzpicture}
\begin{scope}
\clip (5,0) circle (2.5cm);
\draw (3.4,1.85) circle (.2cm);
\draw (3.4,1.85) circle (.4cm);
\draw (2.5,.5) circle (.35cm);
\draw (2.5,-.5) circle (.35cm);
\draw (3.4,-1.85) circle (.35cm);
\draw (6.5,1.85) circle (.35cm);
\draw (7.5,.5) circle (.35cm);
\draw (7.5,-.5) circle (.35cm);
\end{scope} 
\draw[dashed] (5,0) circle (2.5cm);
\draw  (2.8,1.18743) to[out=0,in=90]  (5,0);
\draw  (2.8,-1.18743) to[out=0,in=-90]  (5,0);
\draw  (4.,2.29129) to[out=-10,in=90]  (5,0);
\draw  (4.,-2.29129) to[out=10,in=-90]  (5,0);
\draw (5,0) node {$\bullet$};
\begin{scope}
\clip (12.5,0) circle (2.5cm);
\draw (10.9,1.85) circle (.2cm);
\draw (10.9,1.85) circle (.4cm);
\draw (10,.5) circle (.35cm);
\draw (10,-.5) circle (.35cm);
\draw (10.9,-1.85) circle (.35cm);
\draw (14,1.85) circle (.35cm);
\draw (15,.5) circle (.35cm);
\draw (15,-.5) circle (.35cm);
\end{scope} 
\draw[dashed] (12.5,0) circle (2.5cm);
\draw (13.25,0) circle (.75cm);
\draw  (10.3,1.18743) to[out=0,in=90]  (12.5,0);
\draw  (10.3,-1.18743) to[out=0,in=-90]  (12.5,0);
\draw (12.5,0) node {$\bullet$};
\end{tikzpicture}
\vskip .5cm
\caption{The left figure displays a typical diagram contributing to the  correlator $ \langle W {\rm tr} \varphi^{4} \rangle$. The dashed line stays for the Wilson loop. Solid lines 
denote gluon/scalar propagators.  The correlator counts graphs involving a single quartic vertex interaction with all four legs ending on the Wilson loop. The counting discards 
 diagrams of the type in the right figure where a pair of legs starting from the quartic vertex contract among themselves. }
\label{figure1}
\end{figure}
\end{center}
In doing so, we should remember that the chiral primary operators involved in the correlator are traceless, so  the diagrams in Fig.\ref{figure} and Fig.\ref{figure1}
with loops starting and ending on the same four-point vertex should be discarded.  Taking $\epsilon_1=\epsilon_2=4\pi $  (\ref{lll}) reads 
  \be
\lambda=  \lambda_{\rm YM} 
 \ee   
 The correlators $ \langle W {\rm tr} \varphi^{J} \rangle$  can be extracted from the
  small  $g_4$-expansion  of (\ref{wilsonfin})  
   \be
  W  = W_0- g_4\, W_1+\ldots 
 \ee
   with  
   \bea
   W_0 &=&   {2\,  I_1 \left( \sqrt{  \lambda_{\rm YM} } \right)  \over\sqrt{  \lambda_{\rm YM}  }} =1+{ \lambda_{\rm YM} \over  4 \, 2!}+ {2 \lambda_{\rm YM} ^2\over 4^2 4!} + {5 \lambda_{\rm YM} ^3\over 4^3 6!}
   +\ldots \nn\\
   W_1 &=&  12\,  \lambda_{\rm YM} ^2\, I_4 \left(\sqrt{  \lambda_{\rm YM}  } \right)  + 48 \,  \lambda_{\rm YM} ^{3\over 2} \, I_3 \left(\sqrt{  \lambda_{\rm YM}  } \right) 
   =      {8  \lambda_{\rm YM} ^3\over 4 \,2! }+ \frac{36 \,  \lambda_{\rm YM} ^4  }{4^2 6!} + \frac{144  \lambda_{\rm YM} ^5  }{4^3 6! } \,+\ldots \label{w0w1} 
    \eea
   The integer coefficients in the numerators of the right hand side expansions (\ref{w0w1})  count the number of diagrams with zero and one 
   $g_4$-vertices in Fig.\ref{figure}.  The insertion of ${\rm tr}\varphi^{2}$ inside the Wilson loop correlator can be obtained from $W_0$
   after taking a derivative with respect to $\lambda$. Indeed using (\ref{ws4}) and (\ref{v2}) one finds
  \be
 \langle {\rm tr} e^{\cal C} \,{\rm tr} \varphi^{2} \rangle  = 2 \lambda_{\rm YM} ^2\frac{\partial}{\partial \lambda_{\rm YM} } W_0=2\, \lambda_{\rm YM} \, I_2\left(\sqrt{ \lambda_{\rm YM} }\right)
\ee
and it agrees with the result in \cite{Semenoff:2001xp}.
The result for $  \langle {\rm tr} e^{\cal C} \,{\rm tr} \varphi^{4} \rangle $ follows from $W_1$ after subtracting the contributions of the diagrams involving
four-point vertices with self contracted legs. These diagrams are counted by
\be
8\,  \lambda_{\rm YM} ^3\, \frac{\partial}{\partial \lambda_{\rm YM} } W_0=8\, \lambda_{\rm YM} \, I_2\left(\sqrt{ \lambda_{\rm YM} }\right)
={8  \lambda_{\rm YM} ^3\over 4 \,2! }+ \frac{32 \,  \lambda_{\rm YM} ^4  }{4^2 6!} + \frac{128  \lambda_{\rm YM} ^5  }{4^3 6! } \,+\ldots
\ee
 Subtracting this contribution from $W_1$ one finds 
\be
 \langle {\rm tr} e^{\cal C} \,{\rm tr} \varphi^{4} \rangle={  \lambda_{\rm YM} ^2 \over 4} \, I_4(\sqrt{ \lambda_{\rm YM} }) =
   \frac{4 \,  \lambda_{\rm YM} ^4  }{4^2 6!} + \frac{16  \lambda_{\rm YM} ^5  }{4^3 6! } \,+\ldots  
\label{trfi4}
\ee
in agreement with  \cite{Semenoff:2001xp}.  Our derivation,  based on localization, can be considered as a proof of the
 formulae derived in \cite{Semenoff:2001xp}  under the assumptions that correlators   $ \langle {\rm tr} e^{\cal C} \,{\rm tr} \varphi^{J} \rangle$ are not corrected by graphs  involving internal vertex interactions. 
The results provide exact (to all orders in perturbation theory at large N) formulae for correlators involving Wilson loops and chiral primary operators generalizing the  Wilson loop results obtained
   in \cite{Pestun:2007rz}  using localization  and in \cite{Drukker:1999zq,Erickson:2000af,Drukker:2000rr} from perturbation theory.

\section{ AGT duality:  chiral correlators vs integrals of motion  }

\subsection{The CFT side}

The AGT correspondence \cite{Alday:2009aq} relates ${\cal N}=2$ supersymmetric gauge theories in four dimensions to
 two dimensional CFTs.  According to this correspondence the instanton partition function for
 the SU(2) gauge theory with four fundamentals is related to four-point  correlators
 in Liouville theory. In this section we study the CFT side of the duality. In the next section we will study the gauge side of it and show that the chiral correlators $\langle {\rm tr} \tilde \varphi^n \rangle$
in the gauge theory are reproduced  by the same four-point correlators
in Liouville theory with the insertion of  the integrals of motion $I_n$ introduced in \cite{Alba:2010qc}.

 \subsubsection{The conformal field theory}

 Here we follow \cite{Alba:2010qc}. We refer the reader to this paper for details and further references.
 The symmetry algebra of Liouville theory is the tensor product of a Virasoro and a Heisenberg algebra with commutation relations
 \bea
 \left[ L_m,L_n \right] &=&(m-n)\, L_{m+n}+{c\over 12}(m^3-m)\,\delta_{m+n,0} \nn\\
 \left[ a_m,a_n \right] &=& {m\over 2} \delta_{m+n}  \qquad \left[ L_m ,a_n\right]=0
\eea
The central charge c is parametrized by
\be
c=1+6 \, Q^2    \qquad {\rm where}  \qquad Q=b+{1\over b}
\ee
The primary fields $V_\alpha$ are defined as
\be
V_\alpha(z) = {\cal V}^{\rm vir}_\alpha (z)  \, {\cal V}^{\rm heis}_\alpha (z) \,
\ee
with ${\cal V}^{\rm vir}_\alpha$ a primary field of the Virasoro algebra with dimension
$\Delta(\alpha)=\alpha(Q-\alpha)$ and
\be
 {\cal V}^{\rm heis}_\alpha (z) =e^{2 i (\alpha-Q) \sum_{n<0} {a_n\over n} z^{-n} } \,e^{2 i \alpha \sum_{n>0} {a_n\over n} z^{-n} }
 \ee
 The commutation relations of the field $V_\alpha $ and the generators $L_m$, $a_n$ are
 \bea
  \left[ L_m, V_\alpha (z) \right] &=&
{\cal V}^{\rm heis}_\alpha (z)
\left( z^{m+1}\partial_z+(m+1) \Delta(\alpha)\, z^m\right) \, {\cal V}^{\rm vir}_\alpha (z) \nn\\
   \left[ a_n, V_\alpha(z) \right] &=&
\left\{
\begin{array}{cc}
 {\rm i} \, (Q-\alpha) \, z^n \,   V_\alpha (z)   &   {\rm for} \qquad n>0 \\
  - {\rm i} \, \alpha \, z^n \,   V_\alpha (z)   &   {\rm for} \qquad n<0 \\
\end{array}
\right.
   \label{commrel}
 \eea
  The Fock space is obtained by acting with $L_n,a_n$ with $n<0$ on a vacuum $|0\rangle$ defined by
 \be
 L_m |0\rangle=a_n |0\rangle=0    \qquad {\rm for} \qquad m \geq -1,\quad n>0
 \ee
 Primary states $|\alpha\rangle$ and $\langle \alpha | $ are obtained by acting on the vacuum with the primary fields at zero and infinity respectively
 \be
 |\alpha\rangle =V_\alpha (0) \, |0\rangle     \qquad ~~~~~~~~~~~~~~~~~
 \langle \alpha | = \lim_{z\rightarrow\infty}z^{2\Delta(\alpha)}\langle 0 | \,V_\alpha (z)
 \ee
Any correlator of the composite fields $V_{\alpha}$ factorizes into the product of a Heisenberg and a Virasoro part. The Heisenberg part, which are just free bosons, is easy to
compute and it reads
\be
 \langle   {\cal V}^{\rm heis}_{\alpha_1} (z_1)  \,  {\cal V}^{\rm heis}_{\alpha_2} (z_2)  \,  {\cal V}^{\rm heis}_{\alpha_3} (z_3)  \,
   {\cal V}^{\rm heis}_{\alpha_4} (z_4)  \,  \rangle=\left(1-\frac{z_3}{z_2} \right)^{2\alpha_2(Q-\alpha_3)}
   \label{4heis}
 \ee
 Consider the remaining Virasoro part of the four-point correlator
 \be
 {\cal G}_{\rm vir}\left(\alpha_i,\alpha | z_i \right)=z_1^{{2\Delta(\alpha_1)}}\langle\langle   {\cal V}^{\rm vir}_{\alpha_1} (z_1)  \,  {\cal V}^{\rm vir}_{\alpha_2} (z_2)  \,  {\cal V}^{\rm vir}_{\alpha_3} (z_3)  \,
   {\cal V}^{\rm vir}_{\alpha_4} (z_4)  \, \rangle\rangle_{\alpha}
 \ee
where by $\langle\langle ~\rangle\rangle_{\alpha}$ we denote the four-point conformal block involving the exchange of a state of conformal dimension
   \be
   \Delta=\Delta(\alpha)
   \ee
and the factor $z_1^{{2\Delta(\alpha_1)}}$ is included to guarantee a finite limit at $z_1 \rightarrow \infty $.
The fact that the 4-point correlator  depends non-trivially only on the cross ratio follows from conformal invariance, so without loss of generality  we can fix three points, let us say $z_1=\infty$, $z_2=1$ and $z_4=0$ and denote the resulting function of a single variable $z\equiv z_3$
as $ {\cal G}_{vir}(\alpha_i,\alpha|z)$ or simply as $ {\cal G}_{vir}$
if it is clear from the context, what are the argument and the parameters.
Derivatives $ \partial_{z_i} {\cal G}_{\rm vir}$ of the correlator can be also written in terms of derivative with respect to $z$. For the choice above one finds
\cite{Fucito:2013fba}
  \bea
 \partial_{z_1}{\cal G}_{\rm vir}&=& 0 \qquad~~~\qquad   \partial_{z_2} {\cal G}_{\rm vir}=(-z \partial_z
+2 \Delta_1-\delta)\,{\cal G}_{vir}    \nn\\
  \partial_{z_3} {\cal G}_{\rm vir} &=& \partial_z \,{\cal G}_{vir}  \quad \qquad
  \partial_{z_4}{\cal G}_{\rm vir}=((z-1) \partial_z+\delta-2 \Delta_1)\,
{\cal G}_{vir}
 \eea
  with $\delta=\sum_{i=1}^4 \Delta_i$ and $\Delta_i=\Delta(\alpha_i)$.
 Including also the contribution of the Heisenberg sector  one finds the conformal block
 \bea
 {\cal G}(\alpha_i,\alpha|z)&\equiv &\langle\langle \alpha_1 | V_{\alpha_2}(1) \, V_{\alpha_3}(z) \, |\alpha_4 \rangle\rangle_{\alpha}
 = (1-z)^{2\alpha_2(Q-\alpha_3)}{\cal G}_{vir}(\alpha_i,\alpha|z)
\label{confblock}
 \eea
    The ``physical"  correlator can be written
   as the integral of the modulus square of the conformal block (\ref{confblock})
   \be
   G(\alpha_i|z)=\int {d\alpha \over 2\pi} \, C_{\alpha_1 \alpha_2 \alpha}\, C_{\alpha\alpha_3 \alpha_4}\,
   |{\cal G}(\alpha_i,\alpha|z) |^2
\label{fullcorr}
   \ee
   where $C_{\alpha_1\alpha_2\alpha}$ are the Liouville structure
   constants \cite{Zamolodchikov:1995aa,Dorn:1994xn}

\subsubsection{Integrals of motion  }
\label{IM}

 Being an integrable system, the Liouville  theory admits the existence of an infinite set of mutually commuting operators or integrals of motion. Explicitly
the first three such integrals are  \cite{Alba:2010qc}
\bea
 && I_2 = L_0-\ft{c}{24}+2 \sum_{k=1}^\infty a_{-k}\, a_k\nn\\
  && I_3 =   \sum_{k=-\infty,k\ne 0}^{\infty} a_{-k}\, L_k+ 2\, \ii\, Q \sum_{k=1}^\infty k\, a_{-k}\, a_k+
  \ft13 \sum_{i+j+k=0} a_{i}\,a_j\, a_{k}  \\
  && I_4 = 2  \sum_{k=1 }^\infty L_{-k}\, L_k+L_0^2-\ft{c+2}{12}L_0+6 \sum_{k=-\infty,k\ne 0}^{\infty}\sum_{i+j=k} L_{-k}\, a_i\, a_j
+12(L_0-\ft{c}{24})\sum_{k=1}^\infty  a_{-k}\, a_k\nn\\
  &&+6 \ii Q \sum_{k=-\infty,k\ne 0}^{\infty}|k|  a_{-k}\, L_k +2(1-5\, Q^2)\sum_{k=1}^\infty k^2\, a_{-k}\, a_k+
  6 \ii Q\sum_{i+j+k=0} |k| a_{i}\,a_j\, a_{k}\nn\\
&& +\sum_{i+j+k+l=0} :\, a_{i}\,a_j\, a_k \,a_{l} \,: \nn
 \label{intmotion}
\eea
 We can insert these operators inside the four-point correlators and the corresponding conformal blocks.  We define
 \be
 {\cal G}_n(\alpha_i,\alpha|z)=\langle\langle \alpha_1 | V_{\alpha_2}(1) \,I_n\,  V_{\alpha_3}(z) \, |\alpha_4 \rangle\rangle_{\alpha}
 \ee
 To compute  ${\cal G}_n$, we can use the commutation relations (\ref{commrel}) to bring creator and annihilator operators to the left and right sides of the correlation respectively. For instance \footnote{One
should be careful to take into account that the commutators $[L_n,V_{\alpha}]$ produce derivatives only of the Virasoro part of the
 composite field $V_{\alpha}$.
}
  {\small
 \bea
  {\cal G}_2 &=& \langle\langle \alpha_1 | V_{\alpha_2}(1) \,  [L_0,V_{\alpha_3}(z)] \, |\alpha_4 \rangle\rangle_{\alpha} +2 \sum_{k=1}^\infty
  \langle\langle \alpha_1 | [V_{\alpha_2}(1),a_{-k}] \,  [a_k,V_{\alpha_3}(z)] \, |\alpha_4 \rangle\rangle_{\alpha} +(\Delta_4-\ft{c}{24} ){\cal G}\nn\\
  &=& \left( z\partial_z+ {2\alpha_2(Q-\alpha_3) z\over 1-z}+\Delta_3+\Delta_4-\ft{c}{24}- {2\alpha_2(Q-\alpha_3) z\over 1-z} \right) {\cal G}\nn\\
&=&\left( z\partial_z+\Delta_3+\Delta_4-\ft{c}{24}\right)
{\cal G}
 \eea
 }
 Similarly for ${\cal G}_3$ one finds
 {\small
 \bea
 {\cal G}_3
  &=&   \ii \sum_{k=1}^\infty     z^k  \left[ z (Q+\alpha_2-\alpha_3)  \left(  \partial_{z} +{2 \alpha_2 (Q-\alpha_3)\over 1-z }\right)
  +(k+1) \alpha_2 \Delta_3 +(k-1)(Q-\alpha_3)   \Delta_2 \right] {\cal G} \nn\\
  && \ii \left[ -2Q \alpha_2(Q-\alpha_3)  \sum_{k=1}^\infty     z^k
   +\alpha_2 (Q-\alpha_3) (Q+\alpha_2-\alpha_3)  \sum_{i,j=1}^\infty z^{i+j}     \right] {\cal G} \nn\\
   &=& {\ii \, z \over 1-z} \left[  (Q+\alpha_2-\alpha_3)  \, z \partial_z+(Q-\alpha_3)(\Delta_2+\Delta_3+\Delta_4-\Delta_1)-2\alpha_2(Q-\alpha_3)^2) \right]
{\cal G}
 \eea
 }
   Proceeding in this way one can write ${\cal G}_n$ in terms of
 ${\cal G}$ and their $z$-derivatives, or as differential operators acting on ${\cal G}$.   We write
  \be
{\cal G}_n(\alpha_i,\alpha|z)=  {\cal L}_n \, {\cal G}(\alpha_i,\alpha|z)      \label{gndef}
 \ee
 with ${\cal L}_n$ a differential operator on the variable $z$. For the first few terms one finds
 \bea
 {\cal L}_2 &=&   z\partial_z+\Delta_3+\Delta_4-\ft{c}{24}  \label{ldiff}\\
 {\cal L}_3 &=&  {i z\over 1-z} \left[ (Q+\alpha_2-\alpha_3) \, z\, \partial_z +(Q-\alpha_3)(\Delta_2+\Delta_3+\Delta_4-\Delta_1)
  -2 \alpha_2(Q-\alpha_3)^2  \right]\nn
 \eea
 We have also computed an analogous formula for ${\cal L}_4$, but it is too lengthy to be reproduced here. Instead, later we will present its gauge theory counterpart, which is less cumbersome (see eq. (\ref{recurr})).

\subsection{The gauge/CFT dictionary}

The AGT correspondence relates the four-point conformal block of the Liouville theory to the partition function of  the ${\cal N}=2$ supersymmetric $SU(2)$ gauge theory with four fundamentals.  The gauge coupling parameter $q=e^{2\pi {\rm i} \tau}$  is identified with the harmonic ratio $z$ parametrizing the positions of vertex insertions. The gauge theory masses $m_u, \bar m_{2+u}$  are related to the conformal dimensions of the vertex insertions in the CFT.
To achieve a full symmetry with respect to the exchange of the four masses we make
the replacements  $m_3\to \bar{m}_3+\epsilon $, $m_4\to \bar{m}_4+\epsilon $.
The vacuum expectation value $a$ for the scalar field at infinity  parametrizes the dimension of the exchanged state.
The squeezing parameter $\epsilon_1/\epsilon_2$ characterizing the $\Omega$ gravitational background parametrizes the central charge of the CFT. The full dictionary
is given by \cite{Alday:2009aq}
\bea
\alpha_1 &=&  \ft{\epsilon}{2}+\ft12(\bar{m}_1-\bar{m}_2)
 \qquad   \alpha_2 =  -\ft12 (\bar{m}_1+\bar{m}_2 ) \nn\\
  \alpha_3 &=& \epsilon+\ft12 ( \bar m_3+\bar m_4)  \qquad \alpha_4= \ft{\epsilon}{2}+  \ft12( \bar m_3- \bar m_4)
 \nn\\
  \alpha &=&   \ft{\epsilon}{2}+  a  \qquad \epsilon=\epsilon_1+\epsilon_2=Q\qquad
  \epsilon_1= b \qquad  \epsilon_2=b^{-1}  \qquad z=q     \label{agt}
   \eea
  The instanton partition function of the gauge theory on $
\R^4$ is  related to the conformal block ${\cal G}(\alpha_i,\alpha|q)$ via
 \be
 Z^{U(2)}_{\rm tree+inst}  (a,m_i,q) = q^{-a^2  }  \, Z^{U(2)}_{\rm inst} (a,m_i,q) =q^{ -{Q^2 \over 4} +\Delta_3 +\Delta_4} \,
 {\cal G}(\alpha,\alpha_i,q)
\label{ZU2}
 \ee
  with $Z_{\rm inst}\sim 1$ and ${\cal G}\sim
   q^{\Delta-\Delta_3 -\Delta_4} $ for small $q$.
 On the other hand,  the Virasoro conformal block is related to the SU(2) partition function  via
  \be
{\cal G}_{\rm vir}(\alpha,\alpha_i, q)   = q^{ \Delta-\Delta_3 -\Delta_4}  \,  Z^{SU(2)}_{\rm inst} (a,m_i q) = q^{ \Delta-\Delta_3 -\Delta_4}  \,  (1-q)^{-2\alpha_2(Q-\alpha_3)} \,
Z_{\rm inst}^{U(2)} (a,m_i q)
 \label{ZSU2} \ee
  with  the extra factor canceling the $U(1)$  contribution (\ref{4heis})  arising from the Heisenberg CFT field.
The full four-point correlator (\ref{fullcorr})
 is then identified with the gauge partition function on the sphere via   \be
   G(\alpha_i,q) =  Z^{U(2)}_{S^4} (m_i, q)
   \ee

 \subsection{  The gauge theory side }

 It is known \cite{Nekrasov:2009rc,Bonelli:2009zp,Bonelli:2014iza} that the integrals of motion (\ref{intmotion}) can be put in relation to the chiral correlators $\left\langle {\rm tr}\, \tilde{\varphi}^J \right\rangle$.
In this section we translate (\ref{ldiff}) in terms of the gauge theory variables to find that the chiral correlators in the undeformed $U(2)$ gauge theory
can be expressed  in terms of q-derivatives of the partition function $Z$. This leads to chiral ring type relations valid at all-instanton orders for a finite $\Omega$-background.
The results will be checked against a microscopic instanton computation and in the so called Nekrasov-Shatashvili limit,  where one of the two parameters of the $\Omega$-background goes  to zero, with the analog of the Seiberg-Witten curve  obtained in \cite{Poghossian:2010pn,Fucito:2011pn}.

\subsubsection{Chiral relations: $\epsilon_1,\epsilon_2$ finite}

 The results (\ref{gndef}) and (\ref{ldiff}) can be translated into chiral correlators using the identification
  \bea
 \langle  {\rm tr} \tilde{\varphi}^2 \rangle &=&  -2\,  \frac{{\cal G}_2(q)}{{\cal G}}- \ft{1}{12}          \nn\\
    \langle  {\rm tr} \tilde{\varphi}^3 \rangle &=& 6 \,\ii \,  \frac{{\cal G}_3(q)}{{\cal G}}\\
\langle  {\rm tr} \tilde{\varphi}^4 \rangle &=&2 h^4\frac{{\cal G}_4(q)}{{\cal G}}-\frac{h^2}{4}\langle  {\rm tr} \tilde{\varphi}^2 \rangle+\frac{\epsilon^2(h^2+\epsilon^2)}{8}\nn
 \eea
 where
 \be
   \epsilon= \epsilon_1+\epsilon_2 \qquad ~~~~~~~~    h^2= \epsilon_1 \epsilon_2
   \ee
Using the AGT dictionary ( \ref{agt}), (\ref{ZU2}) leads to
   {\small
    \bea
     \langle  {\rm tr} \tilde{\varphi}^2 \rangle &=&   - 2 \, h^2 \,   {q\,\partial_q Z  \over   Z } \nn\\
    \langle  {\rm tr} \tilde{\varphi}^3 \rangle &=& \frac{ 3 \, q }{1-q}
      \, \left ( - h^2 \,  {M}_{1}   {q\,\partial_q Z \over   Z  }  +{M}_{3}  \right )\nn\\
       \langle  {\rm tr} \tilde{\varphi}^4 \rangle &=&
       \frac{  2 \, q  }{{\left ( 1-q\right ) }^{2}}    \, \left ( 2  \, \epsilon\, {M}_{3} +2 M_4  +2 \, q \, \left ( {M}_{1} \, {M}_{3} -M_4\right )   +
     h^4  \,q(1-q^2)  { \partial^2_q Z  \over   Z  }   \right. \nn\\
     && \left.   +h^2 \,\left[ h^2-2 q (\epsilon M_1+M_2)+q^2(-h^2+2 M_2-2 M_1^2)  \right]{ \partial_q Z  \over   Z  } \right) \label{recurr}
           \eea
   }
with $Z=Z^{U(2)}_{\rm one-loop} Z^{U(2)}_{\rm inst+tree}$ and
   \be
   M_1=-\sum_{i=1}^4 \bar{m}_i  \quad   M_2=\sum_{i<j}^4 \bar{m}_i \bar{m}_j \quad
    M_3=-\sum_{i < j< k}^4 \bar{m}_i \bar{m}_j \bar{m}_k  \quad M_4=\,
\bar{m}_1 \bar{m}_2 \bar{m}_3 \bar{m}_4
   \ee
   We notice that the last two equations of (\ref{recurr}) can be rewritten in the form
 \begin{eqnarray}
  \langle  {\rm tr} \tilde{\varphi}^3 \rangle &=& {3\,q\over 1-q} \left( \ft12 \,   \langle  {\rm tr} \tilde{\varphi}^2 \rangle\, M_1+  M_3\right) \label{recurr22} \\
  \langle  {\rm tr} \tilde{\varphi}^4 \rangle &=& {(1+q)\,   \over 2(1-q)}  \langle  {\rm tr} \tilde{\varphi}^2 \rangle^2
  + {2q\over (1-q)^2}
  \left(   M_2(1-q) +q M_1^2 +\epsilon M_1 \right)  \langle  {\rm tr} \tilde{\varphi}^2 \rangle
  \nn\\
 &&
 -h^2 {1+q\over 1-q}\, q\partial_q \langle  {\rm tr} \tilde{\varphi}^2 \rangle
   + {4q\over (1-q)^2}
  \left(   M_4(1-q) +q M_1 M_3 +\epsilon M_3 \right) \nn
\end{eqnarray}
which shows that in a finite $\Omega$-background, chiral correlators can be written in terms of  both $\langle  {\rm tr} \tilde{\varphi}^2 \rangle$ and its derivatives.
The chiral ring equations (\ref{recurr22}) generalize those found in \cite{Cachazo:2002ry} to the case of finite $\epsilon_1,\epsilon_2$.
The relations (\ref{recurr}) can be checked against a microscopic instanton computation. Using (\ref{zinst0}-\ref{phij})
one finds
{\small
\bea
 \left\langle {\rm tr}\, \tilde{\varphi}^2 \right\rangle &=&  2 \, {a}^{2}+\frac{ 2 \, q}{4\,a^2-\epsilon^2}
  \, \left ( 2 \, {a}^{4}+{a}^{2} \, \left ( -\epsilon \, {M}_{1}+2 \, {M}_{2}\right ) -\epsilon \, {M}_{3}+2 \, {M}_{4}\right )  +\ldots \nn\\
  \left\langle {\rm tr}\,\tilde{\varphi}^3 \right\rangle &=& q \, \left ( 3 \, {a}^{2} \, {M}_{1}+3 \, {M}_{3}\right ) +\ldots \nn\\
  \left\langle {\rm tr}\, \tilde{\varphi}^4 \right\rangle &=&   2 \, {a}^{4}+\frac{2q}{4 \, {a}^{2}-{\epsilon}^{2}}
  \, \left[ 2 \, {a}^{2} \, \left ( 6 \, {a}^{2}-{h}^{2}-{\epsilon}^{2}\right )  \, \left ( {a}^{2}+{M}_{2}\right ) + \epsilon \, \left ( 6 \, {a}^{2}+{h}^{2}-2 \, {\epsilon}^{2}\right )  \, \left ( {a}^{2} \, {M}_{1}+{M}_{3}\right )
  \right. \nn\\
&&\left.   +2 \, \left ( 6 \, {a}^{2}-{h}^{2}+{\epsilon}^{2}\right )  \, {M}_{4}\right ]  \label{reschir}
\eea
      }
where for the sake of conciseness we have omitted higher powers in $q$ (although we have carried out computations up to $q^4$).
For the gauge theory partition function  one finds
  \bea
  Z_{\rm inst+tree} &=&  q^{-{a^2\over h^2} }
  \left(   1-{q\over h^2(4a^2-\epsilon^2)}(2 a^4+a^2(-\epsilon M_1+2 M_2)-\epsilon\, M_3+2\, M_4)+\ldots   \right) \nn\\
   \label{partfunc}
  \eea
The explicit form of $Z_{\rm one-loop}$ is irrelevant to our purposes since it is $\tau$ independent.
It is then not hard to get convinced that  (\ref{reschir}) and (\ref{partfunc})  satisfy the chiral ring relations (\ref{recurr})  to order $q$.  We have checked this
up to order $q^4$.

\subsubsection{Deformed Seiberg-Witten curve:  $h=\epsilon_1 \epsilon_2=0$}

  Chiral relations of the type of (\ref{recurr}) are believed to exist  for all chiral correlators but the explicit form of these relations becomes quickly too complicated.
The situation is dramatically improved in the limit $h \to 0$,  where as we will show
the  chiral correlators can be efficiently extracted from the deformed Seiberg-Witten curve governing the dynamics of the theory.
 We  perform the limit keeping  $\langle  {\rm tr} \tilde{\varphi}^2 \rangle$ or  equivalently ${\cal F}=-h^2 \ln Z_{\rm inst}$   finite.
 For $SU(2)$ gauge theory with four fundamentals the deformed Seiberg-Witten curve is given by the difference equation \cite{Poghossian:2010pn,Fucito:2011pn}
 \be
 -q \, Q(z-\epsilon) \, y(z)\, y(z-\epsilon) +(1+q)\, P(z) y(z) -1=0  \label{swcurve}
 \ee
 with
 \be
 P(z)=z^2-u_1 z+u_2   \qquad Q(z)=1+\sum_{\ell=1}^4  M_\ell \,z^\ell
 \ee
   The chiral correlators can be extracted from the expansion at large $z$ of
   \be
   \partial_z \, \log y(z)= \left \langle {\rm tr} {1\over z-\tilde{\varphi}  } \right\rangle = {2\over z}
    +{\left\langle
  {\rm tr} \tilde{\varphi}    \over z^2 \right\rangle  }
   +{\left\langle
  {\rm tr} \tilde{\varphi}^2  \over z^3 \right\rangle  }+\ldots
  \label{gen}
   \ee
    Writing
    \be
    y(z)=\sum_{i=2}^\infty y_i \,z^{-i} \label{yy}
    \ee
    and using (\ref{gen}) one finds the chiral correlators as functions of the $y_i$'s.
      \bea
   \left \langle {\rm tr} {1\over z-\tilde{\varphi}   } \right\rangle
 & =& \frac{2}{z}+\frac{{y}_{3}}{{z}^{2}}+\frac{-{y}_{3}^{2}+2 \, {y}_{4}}{{z}^{3}}+\frac{{y}_{3}^{3}-3 \, {y}_{3} \, {y}_{4}+3 \, {y}_{5}}{{z}^{4}} \nn\\
  &&+\frac{-{y}_{3}^{4}+4 \, {y}_{3}^{2} \, {y}_{4}-2 \, {y}_{4}^{2}-4 \, {y}_{3} \, {y}_{5}+4 \, {y}_{6}}{{z}^{5}}+\ldots
    \eea
  On the other hand , plugging (\ref{yy}) into  the difference equation (\ref{swcurve})
  one can solve for $y_i$  order by order in $i$ in terms of
    $u_1$, $u_2$.  Moreover imposing $ {\rm tr} \tilde\varphi =0$ requires $y_3=0$ and determines $u_1$ while $u_2$
   can be solved in terms of   $\langle  {\rm tr} \tilde\varphi^2 \rangle$. The results are
   \be
   u_1=-\frac{q (M_1-2 \epsilon)}{1+q}   \qquad u_2=\frac{(-1+q) \langle  {\rm tr} \tilde{\varphi}^2 \rangle+2 q M_2
   +2 q \epsilon( \epsilon-M_1)}{2 (1+q) }
   \ee
   Using these relations one  can relate all correlators to  $\langle  {\rm tr} \tilde{\varphi}^2\rangle$ finding equations of the chiral ring type.
   Explicitly
 {\small
  \bea
  \langle  {\rm tr} \tilde{\varphi}^3 \rangle &=& {3\,q\over 1-q} \left( \ft12 \,   \langle  {\rm tr} \tilde{\varphi}^2 \rangle\, M_1+  M_3\right) \nn \\
  \langle  {\rm tr} \tilde{\varphi}^4 \rangle &=& {(1+q)\,   \over 2(1-q)}  \langle  {\rm tr} \tilde{\varphi}^2 \rangle^2
  + {2q\over (1-q)^2}
  \left(   M_2(1-q) +q M_1^2 +\epsilon M_1 \right)  \langle  {\rm tr} \tilde{\varphi}^2 \rangle
 \label{chirringe} \\
 &&  + {4q\over (1-q)^2}
  \left(   M_4(1-q) +q M_1 M_3 +\epsilon M_3 \right) \nn
    \eea
    }
     Consistently, (\ref{chirringe}) reduces to (\ref{recurr22}) when $h$ is set to zero.
  Similar relations  can be found for chiral correlators involving  higher powers of the scalar field.

\section{AGT duals of Minimal Models}

We saw before in (\ref{epsratio}) that for the Wilson loop to be closed the ratio of the two parameters of the $\Omega$-background
must be the ratio of two integers. These rational $\Omega$-backgrounds has been conjectured to be  AGT dual of the minimal models
  \cite{Santachiara:2010bt,Estienne:2011qk,Bershtein:2014qma,Alkalaev:2014sma,Belavin:2015ria}.
In this section we  collect evidence in favor of this duality.  It is also possible to show that the insertion of a Wilson loop in the gauge partition function
dual to the minimal model correlators is trivial since it contributes only an overall sign as a consequence of the rationality of the $\Omega$-background.

\subsection{The minimal models}
\label{mm}

We start by reviewing the results  for the basic correlators on the minimal models, with the focus on the Ising model.
Let $p$, $q$ be positive co-prime integers. The minimal
model ${\cal M}_{p,q}$ is characterized by its Virasoro central charge
\begin{eqnarray}
c=1-\frac{6(p-q)^2}{pq}.
\end{eqnarray}
There are $(p-1)(q-1)/2$ primary fields denoted as $\phi_{m,n}$,
$m\in \{1,2,\ldots ,p-1\}$, $n\in \{1,2,\ldots ,q-1\}$ with conformal
dimensions
\begin{eqnarray}
\Delta_{m,n}=\alpha_{m,n} (Q-\alpha_{m,n}),
\label{dim}
\end{eqnarray}
where $Q=b+1/b$, $b=i\sqrt{p/q}$ and
\begin{eqnarray}
\alpha_{m,n}=\frac{1-m}{2b} +b\frac{1-n}{2}
\label{alphanm}
\end{eqnarray}
Note  the identification $\phi_{m,n}\equiv \phi_{p-m,q-n}$,
which reflects the symmetry of the dimension (\ref{dim}) with respect to
$\alpha \leftrightarrow Q-\alpha$.

Since a primary field $\phi_{m,n}$ is degenerated at the level
$m n$, any correlation function including this field  satisfies a linear
differential equation of order $m n$. Thus "in principle" all the
correlation functions are calculable. In particular the four point
functions including the field $\phi_{1,2}$ or $\phi_{2,1}$ satisfy a second order
differential equation and can be explicitly expressed in terms of hypergeometric functions.
   An important feature of the physical 4-point function in minimal
models is that only a finite number of primary fields may appear
in the intermediate channels, so the correlator is given by a sum (rather than an integral) of squares of the corresponding conformal
blocks.

 For the sake of exemplification, here and in the following, we will focus on the simplest case $(p,q)=(3,4)$ corresponding  to the Ising model. In this case there are three spinless
primaries
\be
o=\phi_{1,1}\equiv \phi_{2,3} \qquad  \sigma=\phi_{1,2}
\equiv \phi_{2,2} \qquad \varepsilon=\phi_{2,1}\equiv \phi_{1,3}
\label{primaries}\ee
 whose dimensions are
\be
\Delta_o=0,\qquad \Delta_\varepsilon=\frac{1}{16} , \qquad \Delta_\sigma=\frac{1}{2}
\ee
The non-trivial three-point structure constants are
\be
C_{ooo}=C_{\sigma \sigma o}=C_{\varepsilon \varepsilon o}=1 \qquad ~~~~~~C_{\sigma \sigma \varepsilon}=\frac{1}{2} \label{3pt}
\ee
 leading to the Ising fusion rules
 \be
 \sigma \sigma \sim [o]+[\varepsilon]  \qquad \varepsilon \varepsilon \sim [o] \qquad \sigma \varepsilon \sim [\sigma]
\ee
The conformal blocks consistent with these fusion rules are given by  (see e.g. \cite{AlvarezGaume:1989vk})
  \bea
\langle \sigma \sigma \sigma \sigma \rangle_o&=&
\frac{\sqrt{1+\sqrt{1-z}}}{\sqrt{2}\,\, z^{1/8}(1-z)^{1/8} }\qquad   \langle \sigma \sigma \sigma \sigma \rangle_\varepsilon=
\frac{\sqrt{2}\,\, \sqrt{1-\sqrt{1-z}}}{z^{1/8}(1-z)^{1/8}}\nonumber\\
\langle \varepsilon  \varepsilon  \varepsilon  \varepsilon \rangle_o&=&
\frac{z}{1-z}+\frac{1}{z}\qquad     \langle \sigma  \varepsilon  \sigma \varepsilon \rangle_\sigma =
\frac{1-2 z}{\sqrt{z(1-z)}}\nn\\
\langle \varepsilon  \varepsilon  \sigma \sigma \rangle_o &=&
\frac{2-z}{2z^{1/8}\,\, \sqrt{1-z}}\qquad
\langle  \sigma \sigma \varepsilon  \varepsilon  \rangle_o=
\frac{2-z}{2z\,\, \sqrt{1-z}}
\label{cbgaume}
\eea
The subscripts on the l.h.s. label the operators which are exchanged. For the full correlators one finds
\begin{eqnarray}
&&\langle \sigma  \sigma  \sigma  \sigma )\rangle
=\frac{|1+\sqrt{1 - z}|+|1-\sqrt{1 - z}|}{2|z (1 - z)|^{1/4}} \qquad
 \langle \varepsilon  \varepsilon  \varepsilon   \varepsilon  \rangle
=\frac{|1-z+z^2|^2}{|z(1-z)|^2} \nn\\
&&
\langle \sigma \varepsilon   \sigma  \varepsilon  \rangle
=\frac{|1-2z|^2}{|z(1-z)|}
 \qquad
\langle \varepsilon   \varepsilon  \sigma  \sigma  \rangle
=\frac{|1-z/2|^2}{|z|^{1/4}|1-z|}
\label{zfullising}
\end{eqnarray}

\subsection{Degenerated states vs critical masses}

To realize the Ising model we take $(p,q)=(3,4)$ and
  \be
  \epsilon_1=\epsilon_2^{-1}=b=\ii \, \sqrt{3\over 4}
  \ee
The masses, $\bar m_i$, are chosen such that the $\alpha_i$'s in (\ref{agt}) belong to the set
\be
\alpha_i\in \left\{ \alpha_o, \alpha_\sigma,\alpha_\varepsilon \right\} = \left\{  0, - {b\over 2}  , - {1\over 2b}   \right\} \cup \left\{  Q, Q+ {b\over 2}  , Q+ {1\over 2b}   \right\}
\ee
associated to the primary fields (\ref{primaries}). Without any loss in generality we can discard insertions of the identity operator since they lead to lower point correlators.
The two non-trivial possibilities for the vertex insertions at position q correspond to the choice
 \be
\bar m_3+ \bar m_4=
\left\{
\begin{array}{cc}
 \epsilon_1   &~~~~~~~~ {\cal O}_3=\sigma    \\
\epsilon_2 &  ~~~~~~~~ {\cal O}_3=\varepsilon  \\
\end{array}
\right.    \label{m3400}
\ee
 with arbitrary $\bar{m}_1$, $\bar{m}_2$ (see (3.9) in \cite{Fucito:2013fba}). Using (\ref{agt}) one can rewrite (\ref{m3400}) as
  \be
\alpha_3=
\left\{
\begin{array}{cc}
Q+{b \over 2}    &~~~~~~~~ {\cal O}_3=\sigma    \\
Q+{1 \over 2b}   &  ~~~~~~~~ {\cal O}_3=\varepsilon  \\
\end{array}
\right.    \label{m340}
\ee
The gauge theory partition function on $S^4$ is given by an integration over the   vacuum expectation value  $a=a_1=-a_2$. This integration can be formally evaluated with the method of the residues but the
number of poles  is infinite. Still, as shown in  \cite{Fucito:2013fba}, at the  special values of the masses  (\ref{m3400}) the one loop partition function vanishes and only the residue at those special values of $a$
where two single poles collide can contribute. For $SU(2)$ this happens at two points: $a=a_+$ or $a=a_-$
given by
 \bea
   a_+ = -\bar m_4      \qquad   {\rm or} \qquad    a_- &=&   -\bar m_3
 \label{masscrit}
 \eea
   Using (\ref{agt}), (\ref{m340}), (\ref{masscrit})  one finds
    \bea
     \alpha_{+} =\left\{
\begin{array}{c}
\alpha_4 - \frac{b}{2}       \\
\alpha_4 - \frac{1}{2b}    \\
\end{array}
\right.
\quad
  \alpha_{-} =\left\{
\begin{array}{c}
Q-(\alpha_4 + \frac{b}{2} )       \\
Q-(\alpha_4 + \frac{1}{2b} )    \\
\end{array}
\right.
 \quad
 \left.
\begin{array}{c}
  {\cal O}_3=\sigma    \\
  {\cal O}_3=\varepsilon  \\
\end{array}
\right.
  \label{m34}
 \label{alpha4}
 \eea
  Moreover, for $a=a_+$, the instanton partition function receives contributions only from Young tableaux with a single row(column) at $   a_+$ and no boxes at $-   a_+$ for ${\cal O_3}=\sigma(\epsilon)$.
 Similarly, for $a=a_-$ the relevant    single-row(column) tableaux are centered at $   a_-$.
  The whole instanton sum adds to a hypergeometric function  \cite{Fucito:2013fba}
   \bea
 Z^{U(2)}_{\rm inst,+}&=&  {}_2 F_1 \left(^{A_1,A_2}_{~~B} \big| q \right) \label{glio}\\
Z^{U(2)}_{\rm inst,-} &=&  {}_2 F_1 \left(^{1-B+A_1,1-B+A_2}_{~~2-B} \big| q \right)\nn
\label{Zpm}
\eea
with
\bea
{\cal O}_3=\sigma: \quad  A_1&=&(-\alpha_1+\alpha_2+\alpha_+) b=(  a_+ -\bar m_1)\epsilon_1\nonumber\\
A_2&=&(\alpha_1+\alpha_2+\alpha_+ -Q)b=( a_+ -\bar m_2)\epsilon_1   \label{alio1}   \\
B&=&2 \alpha_+b=(2  a_+ +\epsilon) \epsilon_1  \nonumber \\\nn\\
{\cal O}_3=\varepsilon: \quad A_1&=&{-\alpha_1+\alpha_2+\alpha_+ \over b}=\frac{   a_+ -\bar m_1}{\epsilon_1}\nonumber\\
A_2&=&{\alpha_1+\alpha_2+\alpha_+ -Q \over b}=\frac{   a_+ -\bar m_2}{\epsilon_1} \label{alio}\\
B&=&{2 \alpha_+ \over b}=\frac{2   a_+ +\epsilon}{\epsilon_1}  \nonumber
\eea

 \subsubsection{  The partition function and the four-points conformal blocks}

  The critical values  of the masses associated to the four-point conformal blocks of the Ising model  are listed in Table \ref{tableal}.
\begin{table}
\begin{tabular}{|c|c|c|c|}
\hline
 $\langle {\cal O}_1 {\cal O}_2  {\cal O}_3{\cal O}_4\rangle_{ {\cal O}_\pm}$   & $\pm$ &  $(\alpha_1,\alpha_2,\alpha_3,\alpha_4 ; \alpha_\pm)$   &  $(\bar m_1,\bar m_2,\bar m_3,\bar m_4 ; a_\pm)$  \\
 \hline
&&&\\
$\langle \sigma \sigma \sigma \sigma \rangle_\varepsilon$  & +&  $\left(-{b\over 2}, -{b\over 2}, Q+{b\over 2},- {b\over 2} ;-b \right)$ &  $\left(-{\epsilon\over 2}, \epsilon_1+{\epsilon\over 2}, -{\epsilon\over 2},
\epsilon_1+{\epsilon\over 2} ;  -\epsilon_1-{\epsilon\over 2} \right)$\\
$\langle \sigma \sigma \sigma \sigma \rangle_o$ & -&    $\left(-{b\over 2}, -{b\over 2},Q+ {b\over 2}, -{b\over 2} ; Q \right)$ &  $\left(-{\epsilon\over 2}, \epsilon_1+{\epsilon\over 2}, -{\epsilon\over 2}, \epsilon_1+
{\epsilon\over 2} ;  {\epsilon\over 2} \right)$\\
 $\langle  \varepsilon \varepsilon \varepsilon \varepsilon \rangle_o$ & -&    $\left(-{1\over 2b}, -{1\over 2b}, Q+{1\over 2b}, -{1\over 2b} ; Q \right)$ &  $\left(-{\epsilon\over 2}, \epsilon_2+{\epsilon\over 2},
-{\epsilon\over 2}, \epsilon_2+{\epsilon\over 2} ;  {\epsilon\over 2} \right)$\\
 $ \langle \sigma  \varepsilon  \sigma \varepsilon \rangle_\sigma$   & +&  $\left(-{b\over 2}, -{1\over 2b}, Q+{b\over 2}, -{1\over 2b} ;-{b\over 2}-{1\over 2b} \right)$ &  $\left(-\epsilon_1, \epsilon, -\epsilon_2, \epsilon ;
-\epsilon \right)$\\
$ \langle \sigma   \sigma  \varepsilon \varepsilon \rangle_o$   & -&  $\left(-{b\over 2}, -{b\over 2},Q+ {1\over 2b}, -{1\over 2b} ; Q \right)$ &  $\left(-{\epsilon\over 2}, \epsilon_1+{\epsilon\over 2}, -{\epsilon\over 2},
\epsilon_2+{\epsilon\over 2} ;  {\epsilon\over 2} \right)$\\
  $\langle \varepsilon  \varepsilon  \sigma \sigma \rangle_o$  & -&  $\left(-{1\over 2b}, -{1\over 2b},Q+ {b\over 2},- {b\over 2} ; Q \right)$ & $ \left(-{\epsilon\over 2}, \epsilon_2+{\epsilon\over 2},
-{\epsilon\over 2}, \epsilon_1+{\epsilon\over 2} ;  {\epsilon\over 2} \right)$\\
   &&&\\
  \hline
\end{tabular}
\caption{Critical values  of the masses associated to the four-point conformal blocks of the Ising model. ${\cal O}_\pm$ are the operators associated to $\alpha_\pm$.}
 \label{tableal}
\end{table}
Using (\ref{glio}-\ref{alio})  one finds the $U(2)$ instanton partition functions
\bea
   Z^{U(2), {\rm inst} }_{\langle \sigma \sigma \sigma \sigma \rangle_\varepsilon} &=&
{}_2 F_1 \left(\ft34 , \ft54, \ft32,q \right)
=\frac{ \sqrt{2}\,\, \sqrt{1-\sqrt{1-q}}}{   \sqrt{ q(1-q) } }    \\
Z^{U(2), {\rm inst}}_{ \langle \sigma \sigma \sigma \sigma \rangle_o} &=&
{}_2 F_1 (\ft14,\ft34, \ft12; q )=\frac{  \sqrt{1+\sqrt{1-q}}}{ \sqrt{2}\,\,  \sqrt{1-q} }   \label{cft1}  \\
Z^{U(2), {\rm inst}}_{ \langle \varepsilon  \varepsilon  \varepsilon  \varepsilon \rangle_o} &=&
{}_2 F_1  \left(-\ft13,\ft43,-\ft23,q\right)=\frac{1-q+q^2}{   (1-q)^{5\over 3} } \label{cft2}\\
Z^{U(2), {\rm inst}}_{\langle \sigma  \varepsilon  \sigma \varepsilon \rangle_\sigma}  &=&
  {}_2 F_1(-1,-\ft12,-\ft14,q)
 = (1-2 q)   \label{cft4}\\
Z^{U(2), {\rm inst}}_{\langle \sigma   \sigma  \varepsilon \varepsilon \rangle_o} &=&  Z^{U(2), {\rm inst}}_{  \langle \varepsilon  \varepsilon  \sigma \sigma \rangle_o} =
  {}_2 F_1 \left(\ft14,-1,\ft12,q \right)
 =1-\ft12 q
   \label{cft3}
    \eea
 with the subscript indicating the dual CFT correlator.
The Ising conformal blocks (\ref{cbgaume}) are reproduced from (\ref{cft1}-\ref{cft3})  via (\ref{ZSU2})
  \be
 \langle {\cal O}_1 {\cal O}_2  {\cal O}_3{\cal O}_4\rangle_{ {\cal O}_\pm} =
 q^{ \Delta_{\pm} -\Delta_3-\Delta_4} \, Z^{U(2), {\rm inst} }_{ \langle {\cal O}_1 {\cal O}_2  {\cal O}_3{\cal O}_4\rangle_{ {\cal O}_\pm}  }  \times
\left\{
\begin{array}{cc}
 (1-q)^{\alpha_2  b}  & ~~~~~ {\cal O}_3=\sigma  \\
   (1-q)^{\alpha_2\over b}  & ~~~~~ {\cal O}_3=\varepsilon   \\
 \end{array}
\right.
 \label{alio2}
 \ee
  The remaining critical partition functions $Z^{U(2), {\rm inst}}_{ \langle \varepsilon  \varepsilon  \varepsilon  \varepsilon \rangle_{ {\cal O}_+} }$,
$Z^{U(2), {\rm inst}}_{\langle \sigma  \varepsilon  \sigma \varepsilon \rangle_{{\cal O}_-} } $, $Z^{U(2), {\rm inst}}_{\langle \sigma   \sigma  \varepsilon \varepsilon \rangle_{{\cal O}_+} }$, $ Z^{U(2), {\rm inst}}_{  \langle \varepsilon  \varepsilon  \sigma \sigma \rangle_{{\cal O}_+}} $
  have no counterpart in the minimal model side.  Indeed, as we will see in the next section,
  they do not contribute to the gauge partition function on $S^4$ since the one-loop partition function vanishes at the corresponding critical values.

  \subsubsection{The gauge partition function on $S^4$ vs the four-point correlator }

For the critical choice of masses  (\ref{m340}) the $U(2)$ gauge partition function on $S^4$ (\ref{ZS4})  reduces to \cite{Fucito:2013fba}
\be
 Z^{U(2)}_{ \langle {\cal O}_1 {\cal O}_2  {\cal O}_3{\cal O}_4\rangle   }   =    C_+  |q^{ \Delta_{+} -\Delta_3-\Delta_4} \, Z^{U(2)}_{ \langle {\cal O}_1 {\cal O}_2  {\cal O}_3{\cal O}_4\rangle_{ {\cal O}_+}  }  |^2+C_-
  |q^{ \Delta_{-} -\Delta_3-\Delta_4} \, Z^{U(2)}_{ \langle {\cal O}_1 {\cal O}_2  {\cal O}_3{\cal O}_4\rangle_{ {\cal O}_-}  } |^2    \label{zising}
 \ee
where we used  (\ref{ccc}) or equivalently $c=q^{ {Q^2\over 4} -\Delta_3-\Delta_4}$ to rewrite  $c \,q^{-a_\pm^2}= q^{ \Delta_{\pm} -\Delta_3-\Delta_4}$.   The constants $C_\pm$
are defined by
\be
C_\pm ={\rm Res}_{a_\pm} |Z^{\rm one-loop}_{\langle {\cal O}_1 {\cal O}_2  {\cal O}_3{\cal O}_4\rangle_{a}  }    |^2
\ee
 and  can be interpreted in the dual CFT  as the product of the two three-point functions involved
  in the corresponding conformal block
 \be
C_\pm =C_{{\cal O}_1{\cal O}_2{\cal O}_{\pm}}
C_{{\cal O}_{\pm}{\cal O}_3{\cal O}_4}    \label{cpm}
\ee
The ratio between the two is given by \cite{Fucito:2013fba}
  \bea
{C_- \over  C_+ }=   \frac{\gamma(B)\,\gamma(B-1)}{\gamma(A_1) \gamma(A_2)\,\gamma( B-A_1)\gamma(B-A_2)}
\label{cpm0}
   \eea
with $\gamma(x)={\Gamma(x)\over \Gamma(1-x)}$. Plugging  the special values of $\alpha_i$ from Table \ref{tableal} into (\ref{cpm0})  one finds
\be
\begin{array}{c|c c c c c}
      &   \langle \sigma \sigma \sigma \sigma \rangle & \langle \varepsilon  \varepsilon  \varepsilon \varepsilon \rangle &
      \langle \sigma \varepsilon  \sigma \varepsilon \rangle &   \langle \sigma \sigma \varepsilon   \varepsilon \rangle &
      \langle \varepsilon  \varepsilon  \sigma  \sigma \rangle \\
   \hline
  C_-/C_+ &    4  & \infty  & 0  & \infty &  \infty
\end{array}
\ee
 Adopting the natural normalization
$C_{{\cal O}{\cal O} o}=1$ one gets from (\ref{cpm}) the
correct values of the Ising OPE structure constants (\ref{3pt}).
  Thus one finds the gauge theory  partition functions
  \begin{eqnarray}
 &&Z^{U(2)}_{\langle \sigma  \sigma \sigma  \sigma \rangle}
=\frac{|1+\sqrt{1 - q}|+|1-\sqrt{1 - q}|}{2|q|^{1/4} |1 - q|};\nn\\
 &&Z^{U(2)}_{\langle \varepsilon  \varepsilon  \varepsilon \varepsilon \rangle}
=\frac{|1-q+q^2|^2}{|q|^2 |1-q|^{10\over 3} };\nn\\
&&Z^{U(2)}_{\langle \varepsilon  \varepsilon  \sigma  \sigma \rangle}
=\frac{|1-q/2|^2}{|q|^{1/4} }\nn\\
&&Z^{U(2)}_{\langle \sigma \varepsilon  \sigma \varepsilon \rangle}
=\frac{|1-2q|^2}{|q|}
\label{pyscorr}
\end{eqnarray}
  Notice that  all formulae in the right hand side of (\ref{pyscorr}) are well defined in the entire complex plane. In particular, going around $q=1$ the
  two terms in $Z^{U(2)}_{\langle \sigma  \sigma \sigma  \sigma \rangle}$ get exchanged, so the relative coefficient  $C_-/C_+=4$
  is crucial to ensure the single-valueness of the correlator.
 Finally the Ising model correlators  can be read from (\ref{pyscorr}) after factoring out the $U(1)$ contribution
   \be
 \langle {\cal O}_1 {\cal O}_2  {\cal O}_3{\cal O}_4\rangle =
 Z^{U(2)}_{ \langle {\cal O}_1 {\cal O}_2  {\cal O}_3{\cal O}_4\rangle}  \times
\left\{
\begin{array}{cc}
   |1-q|^{2\alpha_2  b}  &  ~~~~~~~~{\cal O}_3=\sigma  \\
   |1-q|^{2\alpha_2\over b}  & ~~~~~~~~ {\cal O}_3=\varepsilon   \\
 \end{array}
\right.
 \label{alio3}
 \ee
 with results in perfect agreement with (\ref{zfullising}).

\vskip 1.5cm
\noindent {\large {\bf Acknowledgments}}
\vskip 0.2cm
The authors want to thank G.Bonelli and A.Tanzini for many interesting discussions.
The research of R.P. is partially supported by a Visiting Professor Fellowship from the University of Roma Tor Vergata,
by the Volkswagen foundation of Germany, by a grant of the Armenian State Council of Science 13-1C278 and
by the Armenian-Russian grant ``Common projects in Fundamental Scientific Research"-2013.

\providecommand{\href}[2]{#2}\begingroup\raggedright\endgroup

\end{document}